\documentclass[aps,prd,floatfix,superscriptaddress,nofootinbib,11pt]{revtex4-2}
\usepackage{amsmath,bm,amssymb,amsthm,mathrsfs,mathtools}
\usepackage{subfigure,color}
\usepackage{ragged2e}
\justifying\let\raggedright\justifying

\allowdisplaybreaks[4]

\usepackage{tikz}
\usetikzlibrary{backgrounds}
\usetikzlibrary{shapes,matrix,trees}
\usetikzlibrary{arrows.meta}
\usetikzlibrary{positioning}			% For "above of=" commands
\usetikzlibrary{calc,through}			% For coordinates
\usetikzlibrary{decorations.pathreplacing}     % For curly braces
\usepackage{pgffor}                                        % For repeating patterns
\usetikzlibrary{decorations.pathmorphing}	% For Feynman Diagrams
\usetikzlibrary{decorations.markings}
\tikzset{
	vector/.style={decorate, decoration={snake}, draw},
	provector/.style={decorate, decoration={snake,amplitude=2.5pt}, draw},
	antivector/.style={decorate, decoration={snake,amplitude=-2.5pt}, draw},
	fermion/.style={draw=black, postaction={decorate},
		decoration={markings,mark=at position .55 with {\arrow[draw=black]{>}}}},
	fermionbar/.style={draw=black, postaction={decorate},
		decoration={markings,mark=at position .55 with {\arrow[draw=black]{<}}}},
	fermionnoarrow/.style={draw=black},
	gluon/.style={decorate, draw=black,
		decoration={coil,amplitude=4pt, segment length=5pt}},
	scalar/.style={dashed,draw=black, postaction={decorate},
		decoration={markings,mark=at position .55 with {\arrow[draw=black]{>}}}},
	scalarbar/.style={dashed,draw=black, postaction={decorate},
		decoration={markings,mark=at position .55 with {\arrow[draw=black]{<}}}},
	scalarnoarrow/.style={dashed,draw=black},
	electron/.style={draw=black, postaction={decorate},
		decoration={markings,mark=at position .55 with {\arrow[draw=black]{>}}}},
	bigvector/.style={decorate, decoration={snake,amplitude=4pt}, draw},
	photon/.style={decorate, draw=black,decoration={snake,amplitude=4pt, segment length=5pt} }
}

\definecolor{ccblue}{rgb}{0.0,0.4,0.8}
\usepackage[]{hyperref}
\hypersetup{  colorlinks=true,
	linkcolor=ccblue,
	urlcolor=ccblue,
	citecolor=ccblue}

\begin{document}
%%title%%
\title{The Neutrinoless Double Beta Decay in the Colored Zee-Babu Model}
	
%%author%%
\author{Shao-Long Chen}
\email[E-mail: ]{chensl@mail.ccnu.edu.cn}
\affiliation{Key Laboratory of Quark and Lepton Physics (MoE) and Institute of Particle Physics, Central China Normal University, Wuhan 430079, China}
\affiliation{Center for High Energy Physics, Peking University, Beijing 100871, China}
\author{Yu-Qi Xiao}
\email[E-mail: ]{xiaoyq@mails.ccnu.edu.cn}
\affiliation{Key Laboratory of Quark and Lepton Physics (MoE) and Institute of Particle Physics, Central China Normal University, Wuhan 430079, China}

%%abstract%%
\begin{abstract}
We study the neutrinoless double beta decay in the colored Zee-Babu model. We consider three cases of the colored Zee-Babu model with a leptoquark and a diquark introduced. The neutrino masses are generated at two-loop level, and the constraints given by tree-level flavor violation processes and muon anomalous magnetic moment $(g-2)_{\mu}$ have been considered. In our numerical analysis, we find that the standard light neutrino exchange contribution can be canceled by new physics contribution under certain assumption and condition, leading to a hidden neutrinoless double beta decay. The condition can be examined comprehensively by future complementary searches with different isotopes.
\end{abstract}
	
\maketitle
	
%%sec1%%    
\section{Introduction}\label{sec1}
It is widely assumed that the tiny masses of neutrinos could be generated radiatively where neutrinos are Majorana particles. The Majorana neutrino mass models at two-loop level have been discussed in many previous works, e.g., Ref.~\cite{Cheng:1980qt,Petcov:1984nz,Zee:1985id,Babu:1988ki,Babu:2010vp,Angel:2013hla,AristizabalSierra:2014wal,Cao:2017xgk}, among which the Zee-Babu model~\cite{Zee:1985id,Babu:1988ki} has attracted much attention. The addition of new particles in loops can bring us rich phenomena. However, whether neutrinos are Majorana particles or Dirac particles still remains unknown. The search for the neutrinoless double beta ($0\nu\beta\beta$) decay is the promising way to get us out of this dilemma.

The $0\nu\beta\beta$ decay can be realized if neutrinos are Majorana particles. If one only consider the standard light neutrino exchange, the inverse half-life has the form
\begin{align}
\left[T_{1/2}^{0\nu\beta\beta}\right]^{-1}=G_{\nu}|\mathcal{M}_{\nu}|^{2}\dfrac{|\langle m_{ee}\rangle|^{2}}{m_{e}^{2}}\,,
\end{align}
where $G_{\nu}$ and $\mathcal{M}_{\nu}$ are the phase space factor (PSF) and nuclear matrix element (NME), $\langle m_{ee}\rangle\equiv\sum_{i}U_{ei}^{2}m_{i}$ is the effective neutrino mass, and $m_{i}~(i=1,2,3)$ are the masses of neutrinos. The most stringent limit on the $0\nu\beta\beta$ decay half-life in $^{136}{\rm{Xe}}$ isotope is $T_{1/2}^{0\nu\beta\beta}>1.07\times10^{26}~{\rm{yrs}}$ given by KamLAND-Zen experiment~\cite{KamLAND-Zen:2016pfg}. They obtained a constraint of $|\langle m_{ee}\rangle|<61-165~{\rm{meV}}$. The GERDA experiment has published their result $T_{1/2}^{0\nu\beta\beta}>1.8\times10^{26}~{\rm{yrs}}$ with isotope $^{76}{\rm{Ge}}$ leading to a similar bound $|\langle m_{ee}\rangle|<79-180~{\rm{meV}}$~\cite{GERDA:2020xhi}. The future $0\nu\beta\beta$ decay experiments CUPID-1T~\cite{CUPID:2022wpt} and LEGEND-1000~\cite{LEGEND:2021bnm} using $^{100}$Mo and $^{76}$Ge isotopes can push the half-life to $10^{27}-10^{28}~{\rm{yrs}}$, leading to a sensitivity to $|\langle m_{ee}\rangle|$ of around $15~{\rm{meV}}$.

In the effective field theory approach, the $0\nu\beta\beta$ decay can be described in terms of effective low-energy operators~\cite{Prezeau:2003xn,delAguila:2012nu,Cirigliano:2017djv,Deppisch:2017ecm,Cirigliano:2018yza,Graf:2022lhj}. The contributions to $0\nu\beta\beta$ decay can be divided into long-range mechanisms~\cite{Pas:1997cp,Pas:1999fc,Helo:2016vsi,Kotila:2021xgw} and short-range mechanisms~\cite{Pas:2000vn,Graf:2018ozy}. The long-range mechanisms involve light neutrinos exchanged between two point-like vertices, which contain the standard light neutrino exchange. The short-range mechanisms involve the dim-9 effective interaction and are mediated by heavy particles. The decomposition of the short-range operators at tree-level has been completely listed in~\cite{Bonnet:2012kh}, and at one-loop level has been discussed in~\cite{Chen:2021rcv}. The short-range mechanisms at the LHC have been considered in~\cite{Helo:2013ika}. An analysis of the standard light neutrino exchange and short-range mechanisms has been given in~\cite{Deppisch:2020ztt}.
	
In this work, we study three cases of the colored Zee-Babu (cZB) model with a leptoquark and a diquark running in the loops. We consider the realization of tiny neutrino mass and contribution to $0\nu\beta\beta$ decay, with the constraints given by tree-level flavor violation processes and $(g-2)_{\mu}$ considered. The B physics anomalies and some other phenomena in the cZB model have been explored in~\cite{Kohda:2012sr,Nomura:2016ask,Chang:2016zll,Guo:2017gxp,Ding:2018jdk,Datta:2019tuj,Saad:2020ihm,Babu:2020hun}. The long-range contributions given by the leptoquarks have been considered extensively in the previous discussion~\cite{Hirsch:1996qy,Hirsch:1996ye,Graf:2022lhj}. We focus on the short-range impact on neutrinoless double beta decay in this model. The simultaneous introduction of the leptoquarks and diquarks in the cZB model can lead to short-range contribution, which can interfere with the standard light neutrino exchange contribution resulting in cancellation.	
	
We organize our paper as follows. In Sec.~\ref{sec2}, we show the three cases in the cZB model and briefly review the constraints on them. In Sec.~\ref{sec3}, we discuss the $0\nu\beta\beta$ decay in each case, including short-range mechanisms and standard light neutrino exchange. Finally, we give our conclusion in Sec.~\ref{sec4}.

%%sec2%%
\section{The Model and Constraints} \label{sec2}
The colored Zee-Babu (cZB) model requires a leptoquark and a diquark to generate the neutrino mass. The diquark is set to be a color sextet under $SU(3)_{C}\times SU(2)_{L}\times U(1)_{Y}$ symmetry where the hypercharge $Y$ is set to be equal to $Q-I_{3}$. The color triplet is not considered because the coupling of fermions and diquark is antisymmetric, which means the vertex $\overline{d^{c}}d\omega$ is zero and cannot contribute to the neutrinoless double beta decay process. The cZB model with a leptoquark (LQ) and a color sextet diquark (DQ) has three cases : 
	
case 1: a singlet LQ $S_{1}\sim(\overline{3},1,1/3)$ and a singlet DQ $\omega_{1}\sim(6,1,-2/3)$, 
	
case 2: a triplet LQ $S_{3}\sim(\overline{3},3,1/3)$ and a singlet DQ $\omega_{1}\sim(6,1,-2/3)$, 
	
case 3: a doublet LQ $\tilde{S}_{2}\sim(3,2,1/6)$ and a triplet DQ $\omega_{3}\sim(6,3,1/3)$.

\noindent The corresponding quantum numbers of the particles in these cases are summarized in Table~\ref{QNofparticles}. 
	
\begin{table}[t]
\centering
       	\begin{tabular}{||c|c||c|c||}
	\hline
	$\text{SM~Particles}$&$\text{Quantum~Number}$&$\text{New Particles}$&$\text{Quantum~Number}$\\
	\hline
	$\Phi=(\phi^{+},\phi^{0})^{T}$&$(1,2,1/2)$&$S_{1}$&$(\overline{3},1,1/3)$\\
	$Q_{L}=(U_{L},D_{L})^{T}$&$(3,2,1/6)$&$\tilde{S}_{2}$&$(3,2,1/6)$\\
	$L_{L}=(\nu_{L},E_{L})^{T}$&$(1,2,-1/2)$&$S_{3}$&$(\overline{3},3,1/3)$\\
	$U_{R}$&$(3,1,+2/3)$&$\omega_{1}$&$(6,1,-2/3)$\\
	$D_{R}$&$(3,1,-1/3)$&$\omega_{3}$&$(6,3,1/3)$\\
	$E_{R}$&$(1,1,-1)$&&\\
	\hline
	\end{tabular}
\caption{The corresponding quantum numbers of the SM particles and new particles under the gauge symmetry $SU(3)_{C}\times SU(2)_{L}\times U(1)_{Y}$. \label{QNofparticles}}
\end{table}
	
In the fermion weak eigenbasis, the Yukawa interactions of these cases can be written as
\begin{align}
-\mathcal{L}_{Y1}\supset\,&
y^{ij}_{1SR}\overline{(U_{R}^{i\alpha})^{c}}E_{R}^{j}S_{1}^{\overline{\alpha}}
+y^{ij}_{1SL}\overline{(Q_{L}^{i\alpha})^{c}}i\sigma^{2} L_{L}^{j}S_{1}^{\overline{\alpha}}  
+z^{ij}_{1SR}\overline{(U_{R}^{i\alpha})^{c}}D_{R}^{j\beta}S_{1}^{*\gamma}\epsilon^{\alpha\beta\gamma}\notag\\
&+z^{ij}_{1SL}\overline{(Q_{L}^{i\alpha})^{c}}i\sigma^{2} Q_{L}^{j\beta}S_{1}^{*\gamma}\epsilon^{\alpha\beta\gamma}
+z^{ij}_{1\omega}\overline{(D_{R}^{i\alpha})^{c}} D_{R}^{j\beta}\omega_{1}^{*\overline{\alpha}\overline{\beta}}+\text{h.c.}\,,\\
-\mathcal{L}_{Y2}\supset\,&
y^{ij}_{3S}\overline{(Q_{L}^{i\alpha})^{c}}i\sigma^{2}(\sigma^{k}S_{3}^{k\overline{\alpha}} )L_{L}^{j} 
+z^{ij}_{3S}\overline{(Q_{L}^{i\alpha})^{c}}i\sigma^{2}(\sigma^{k}S_{3}^{*k\beta})^{T}Q_{L}^{j\gamma}\epsilon^{\alpha\beta\gamma}\notag\\
&+z^{ij}_{1\omega}\overline{(D_{R}^{i\alpha})^{c}} D_{R}^{j\beta}\omega_{1}^{*\overline{\alpha}\overline{\beta}}+\text{h.c.}\,,\\
-\mathcal{L}_{Y3}\supset\,&
y^{ij}_{2S}\overline{D_{R}^{i\alpha}}(\tilde{S}_{2}^{\alpha})^{T} i\sigma^{2}L_{L}^{j} 
+z^{ij}_{3\omega}\overline{(Q_{L}^{i\alpha})^{c}}i\sigma^{2}(\sigma^{k}\omega_{3}^{*k\overline{\alpha}\overline{\beta}})^{T}Q_{L}^{j\beta}+\text{h.c.}\,,
\end{align}
where 
\begin{align}
\sigma^{k}S_{3}^{k}=
\begin{pmatrix}S_{3}^{+1/3}&\sqrt{2}S_{3}^{+4/3}\\
\sqrt{2}S_{3}^{-2/3}&-S_{3}^{+1/3}
\end{pmatrix}\,,\quad
\sigma^{k}\omega_{3}^{k}=
\begin{pmatrix}\omega_{3}^{+1/3}&\sqrt{2}\omega_{3}^{+4/3}\\
\sqrt{2}\omega_{3}^{-2/3}&-\omega_{3}^{+1/3}
\end{pmatrix}\,,
\end{align}
$\psi^{c}$ is the charge conjugate of $\psi$, $i$ and $j$ label fermion generations, and $(\alpha, \beta, \gamma)$ denote the color of $SU(3)_{c}$. The $\sigma^{k}~(k=1,2,3)$ are Pauli matrices, $S_{3}^{k}$ and $\omega_{3}^{k}$ are the componets of $S_{3}$ and $\omega_{3}$ under $SU(2)_{L}$ symmetry, and $\epsilon^{\alpha\beta\gamma}$ is the Levi-Civita symbol. The Yukawa coupling matrices $z_{1SL}$, $z_{1\omega}$, and $z_{3\omega}$ are symmetric, $z_{3S}$ is antisymmetric, while the other coupling matrices are arbitrary~\cite{Davies:1990sc}. To study the phenomenologies, we rewrite the Lagrangian in the mass eigenbasis as,
\begin{align}
-\mathcal{L}_{Y1}\supset\,&
y^{ij}_{1SR}\overline{(U_{R}^{i\alpha})^{c}}E_{R}^{j}S_{1}^{\overline{\alpha}}			
+(V^{*}y_{1SL})^{ij}\overline{(U_{L}^{i\alpha})^{c}}E_{L}^{j}S_{1}^{\overline{\alpha}}\notag\\
&-(y_{1SL}U)^{ij}\overline{(D_{L}^{i\alpha})^{c}}\nu_{L}^{j}S_{1}^{\overline{\alpha}}	
+z^{ij}_{1SR}\overline{(U_{R}^{i\alpha})^{c}}D_{R}^{j\beta}S_{1}^{*\gamma}\epsilon^{\alpha\beta\gamma}\notag\\	&+2(V^{*}z_{1SL})^{ij}(\overline{U_{L}^{i\alpha})^{c}}D_{L}^{j\beta}S_{1}^{*\gamma}\epsilon^{\alpha\beta\gamma}+z^{ij}_{1\omega}\overline{(D_{R}^{i\alpha})^{c}}D_{R}^{j\beta}\omega^{*\overline{\alpha}\overline{\beta}}+\text{h.c.}\,,\\
-\mathcal{L}_{Y2}\supset\,&
\sqrt{2}(V^{*}y_{3S}U)^{ij}\overline{(U_{L}^{i\alpha})^{c}}S_{3}^{-2/3,\overline{\alpha}}\nu_{L}^{j}-(y_{3S}U)^{ij}\overline{(D_{L}^{i\alpha})^{c}}S_{3}^{+1/3,\overline{\alpha}}\nu_{L}^{j}\notag\\
&-(V^{*}y_{3S})^{ij}\overline{(U_{L}^{i\alpha})^{c}}S_{3}^{+1/3,\overline{\alpha}}E_{L}^{j}
-\sqrt{2}y_{3S}^{ij}\overline{(D_{L}^{i\alpha})^{c}}S_{3}^{+4/3,\overline{\alpha}}E_{L}^{j}\notag\\
&-2(z_{3S}V^{\dagger})^{ij}\overline{(D_{L}^{i\alpha})^{c}}S_{3}^{-1/3,\beta}U_{L}^{j\gamma}\epsilon^{\alpha\beta\gamma}+\sqrt{2}(V^{*}z_{3S}V^{\dagger})^{ij}\overline{(U_{L}^{i\alpha})^{c}}S_{3}^{-4/3,\beta}U_{L}^{j\gamma}\epsilon^{\alpha\beta\gamma}\notag\\
&-\sqrt{2}z_{3S}^{ij}\overline{(D_{L}^{i\alpha})^{c}}S_{3}^{+2/3,\beta}D_{L}^{j\gamma}\epsilon^{\alpha\beta\gamma}+z^{ij}_{1\omega}\overline{(D_{R}^{i\alpha})^{c}} D_{R}^{j\beta}\omega_{1}^{*\overline{\alpha}\overline{\beta}}+\text{h.c.}\,,\\
-\mathcal{L}_{Y3}\supset\,&
(y_{2S}U)^{ij}\overline{D_{R}^{i\alpha}}\tilde{S}_{2}^{-1/3,\overline{\alpha}}\nu_{L}^{j}
-y^{ij}_{2S}\overline{D_{R}^{i\alpha}}\tilde{S}_{2}^{+2/3,\overline{\alpha}}E_{L}^{j}\notag\\
&-2(z_{3\omega}V^{\dagger})^{ij}\overline{(D_{L}^{i\alpha})^{c}}\omega_{3}^{-1/3,\overline{\alpha}\overline{\beta}}U_{L}^{j\beta}
+\sqrt{2}(V^{*}z_{3\omega}V^{\dagger})^{ij}\overline{(U_{L}^{i\alpha})^{c}}\omega_{3}^{-4/3,\overline{\alpha}\overline{\beta}}U_{L}^{j\beta}\notag\\
&-\sqrt{2}z_{3\omega}^{ij}\overline{(D_{L}^{i\alpha})^{c}}\omega_{3}^{+2/3,\overline{\alpha}\overline{\beta}}D_{L}^{j\beta}+\text{h.c.}\,.
\end{align}
Here we follow the basis transition $U_{L}^{j}\rightarrow(V^{\dagger})_{jk}U_{L}^{k},~D_{L}^{j}\rightarrow D_{L}^{j},~E_{L}^{j}\rightarrow E_{L}$ and $\nu_{L}^{j}\rightarrow U_{jk}\nu_{L}^{k}$ in~\cite{Dorsner:2016wpm}, where $V$ is the Cabibbo-Kobayashi-Maskawa (CKM) matrix, and $U$ is the Pontecorvo-Maki-Nakagawa-Sakata (PMNS) matrix. The scalar potential involving the leptoquark and the diquark fields contains cubic terms
\begin{align}
V_{1}&\supset\mu_{1}S_{1}^{\overline{\alpha}}S_{1}^{\overline{\beta}}\omega_{1}^{\alpha\beta}+\text{h.c.}\,,\\	
V_{2}&\supset\mu_{2}(S_{3}^{\overline{\alpha}})^{T}S_{3}^{\overline{\beta}}\omega_{1}^{\alpha\beta}+{\text{h.c.}}\notag\\         
&=2\mu_{2}S_{3}^{-2/3,\overline{\alpha}}S_{3}^{+4/3,\overline{\beta}}\omega_{1}^{-2/3,\alpha\beta}+\mu_{2}S_{3}^{+1/3,\overline{\alpha}}S_{3}^{+1/3,\overline{\beta}}\omega_{1}^{-2/3,\alpha\beta}+{\text{h.c.}}\,,\\
V_{3}&\supset\mu_{3}(\tilde{S}_{2}^{\alpha})^{T}(\sigma^{k}\omega_{3}^{k,\alpha\beta})^{*}i\sigma_{2}\tilde{S}_{2}^{\beta}+{\text{h.c.}}\notag\\
&=\mu_{3}\tilde{S}_{2}^{+2/3,\alpha}\omega_{3}^{-1/3,\overline{\alpha}\overline{\beta}}\tilde{S}_{2}^{-1/3,\beta}     
+\sqrt{2}\mu_{3}\tilde{S}_{2}^{-1/3,\alpha}\omega_{3}^{+2/3,\overline{\alpha}\overline{\beta}}\tilde{S}_{2}^{-1/3,\beta}\notag\\
&-\sqrt{2}\mu_{3}\tilde{S}_{2}^{+2/3,\alpha}\omega_{3}^{-4/3,\overline{\alpha}\overline{\beta}}\tilde{S}_{2}^{+2/3,\beta}    
+\mu_{3}\tilde{S}_{2}^{-1/3,\alpha}\omega_{3}^{-1/3,\overline{\alpha}\overline{\beta}}\tilde{S}_{2}^{+2/3,\beta}+{\text{h.c.}}\,.
\end{align}		
For simplicity, we assume the quartic couplings of leptoquark and the SM Higgs doublet $\Phi$ to be vanishing for case 2 and case 3. In addition, the quartic couplings of $\omega_{3}$ and $\Phi$ are also assumed to be negligible. The leptoquark/diquark multiplets are then degenerate in mass, we denote the masses of $S_{1}$, $S_{2}$, $S_{3}$, $\omega_{1}$, and $\omega_{3}$ as $M_{S_{1}}$, $M_{S_{2}}$, $M_{S_{3}}$, $M_{\omega_{1}}$, and $M_{\omega_{3}}$, respectively. In our numeraical analysis, the masses of the leptoquarks and diquarks are taken as $M_{S}\gtrsim 1.5$ TeV and $M_{\omega}\gtrsim 8$ TeV, which accords with the bounds given by the ATLAS and CMS collaboration~\cite{CMS:2018ncu,ATLAS:2020dsk,ATLAS:2019ebv,CMS:2018lab,CMS:2018svy,CMS:2012cyn,CMS:2019gwf}.

%%%sec2A%%%
\subsection{Neutrino masses}
\begin{figure}[t]
\centering
	\begin{tikzpicture}[line width=1pt, scale=2,>=Stealth]
	%%%%
	\path (0:0) coordinate (a0);
	\path (0:0.5) coordinate (a1);
	\path (0:1) coordinate (a2);
	\path (0:1.5) coordinate (a3);
	\path (180:0.5) coordinate (a4);
	\path (180:1) coordinate (a5);
	\path (180:1.5) coordinate (a6);
	\path (90:1) coordinate (a7);
	%%%%
	\draw [scalar] (a7) arc(90:180:1);
	\draw [scalar] (a7) arc(90:0:1);
	\draw [fermion](a6)--(a5);
	\draw [fermion](a4)--(a5);
	\draw [scalar](a7)--(a0);
	\draw [fermion](a0)--(a4);
	\draw [fermion](a0)--(a1);
	\draw [fermion](a1)--(a2);
	\draw [fermion](a3)--(a2);
	%%%%
	\node[below] at (180:1.5) {$\nu_{L}$};
	\node[below] at (0:1.5) {$\nu_{L}$};
	\node[right] at (90:0.4) {$\omega_{1}^{-2/3}$};
	\node[below] at (0:0.25) {$D_{R}$};
	\node[below] at (0:0.75) {$D_{L}$};
	\node[below] at (180:0.25) {$D_{R}$};
	\node[below] at (180:0.75) {$D_{L}$};
	\node[right] at (35:1.05) {$S_{1,3}^{+1/3}$};
	\node at (150:1.25) {$S_{1,3}^{+1/3}$};
	\node at (180:0.5) {$\times$};
	\node at (0:0.5) {$\times$};
	\end{tikzpicture}\quad
	\begin{tikzpicture}[line width=0.8pt, scale=2,>=Stealth]
	%%%%
	\path (0:0) coordinate (a0);
	\path (0:0.5) coordinate (a1);
	\path (0:1) coordinate (a2);
	\path (0:1.5) coordinate (a3);
	\path (180:0.5) coordinate (a4);
	\path (180:1) coordinate (a5);
	\path (180:1.5) coordinate (a6);
	\path (90:1) coordinate (a7);
	%%%%
	\draw [scalar] (a7) arc(90:180:1);
	\draw [scalar] (a7) arc(90:0:1);
	\draw [fermion](a6)--(a5);
	\draw [fermionbar](a4)--(a5);
	\draw [scalar](a7)--(a0);
	\draw [fermionbar](a0)--(a4);
	\draw [fermionbar](a0)--(a1);
	\draw [fermionbar](a1)--(a2);
	\draw [fermion](a3)--(a2);
	%%%%
	\node[below] at (180:1.5) {$\nu_{L}$};
	\node[below] at (0:1.5) {$\nu_{L}$};
	\node[right] at (90:0.4) {$\omega_{3}^{+2/3}$};
	\node[below] at (0:0.25) {$D_{L}$};
	\node[below] at (0:0.75) {$D_{R}$};
	\node[below] at (180:0.25) {$D_{L}$};
	\node[below] at (180:0.75) {$D_{R}$};
	\node[right] at (35:1.05) {$\tilde{S}_{2}^{-1/3}$};
	\node at (150:1.25) {$\tilde{S}_{2}^{-1/3}$};
	\node at (180:0.5) {$\times$};
	\node at (0:0.5) {$\times$};
	\end{tikzpicture}
\caption{The two-loop diagrams which generate tiny neutrino masses in the colored Zee-Babu model. The left diagram corresponds to cases 1 and 2, and the right corresponds to case 3.}\label{neutrino_mass}
\end{figure}
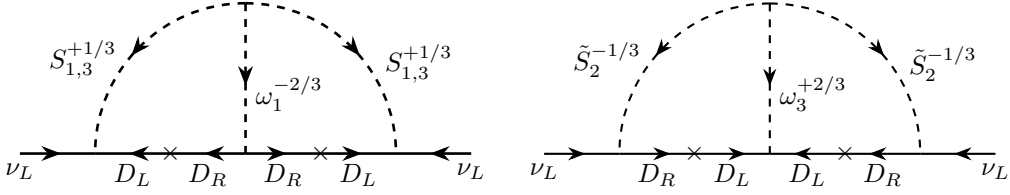
	
In the colored Zee-Babu model with a leptoquark and a diquark, the neutrino masses can be generated at two-loop level with down-type quarks running in the loop as shown in Fig.~\ref{neutrino_mass}. The left Feynman diagram corresponds to case 1 and case 2, and the right one to case 3. The neutrino mass matrix elements in flavor basis take the form~\cite{Kohda:2012sr}
\begin{align}
M_{\nu_{a}}^{kn}=&
24\mu_{a}[y_{bS(L)}^{T}]^{kl}m_{D^l}z_{c\omega}^{lm}m_{D^m} y_{bS(L)}^{mn}\mathcal{I}_{lm}\\
=& 24\mu[y_{S(L)}^{T}]^{kl}m_{D^l}z_{\omega}^{lm}m_{D^m}y_{S(L)}^{mn}\mathcal{I}_{lm}\,,\label{numass}
\end{align}
where $m_{D^l}$ is the mass of the $l$-th generation down-type quark. The superscripts $k,l,m,n=1,2,3$ and the subscripts $a,b,c=1,2,3$ of the couplings are neglected to keep the expression concise. The expression can apply to all three cases. Note that the coupling $y_{S(L)}$ equals $y_{1SL}$ in case 1, $y_{3S}$ in case 2, and $y_{2S}$ in case 3. The $\mathcal{I}_{lm}$ in Eq.~(\ref{numass}) is loop integral
\begin{align}
\mathcal{I}_{lm}=
\int\dfrac{d^{4}k}{(2\pi)^{4}}
\int\dfrac{d^{4}q}{(2\pi)^{4}}
\dfrac{1}{q^{2}-m_{D^{l}}^{2}}
\dfrac{1}{q^{2}-M_{S}^{2}}
\dfrac{1}{k^{2}-M_{S}^{2}}
\dfrac{1}{k^{2}-m_{D^{m}}^{2}}
\dfrac{1}{(k-q)^{2}-M_{\omega}^{2}}\,,
\end{align}
where $M_{S}$ denotes leptoquark $S_{i}$ mass and $M_{\omega}$ is diquark $\omega_{i}$ mass in different cases. The integral can be simplified as~\cite{Babu:2002uu}
\begin{align}
\mathcal{I}_{lm}\simeq \dfrac{1}{(16\pi^{2})^{2}}\dfrac{1}{M_{S}^{2}}\tilde{I}\bigg(\dfrac{M_{\omega}^{2}}{M_{S}^{2}}\bigg)\,,\quad
\tilde{I}(r)&=-\int_{0}^{1}dx\int_{0}^{1-x}dy\dfrac{1}{x+(r-1)y+y^{2}}\ln{\dfrac{y(1-y)}{x+ry}}\,,
\end{align}
with the $\tilde{I}(r)$ can be calculated through the numerical integral way. The neutrino mass matrix can be diagonalized by the PMNS matrix $U$
\begin{align}
U^{T}{M}_{\nu}U=\hat{M}_{\nu}=
\begin{pmatrix}
m_{1}&0&0\\
0&m_{2}&0\\
0&0&m_{3}
\end{pmatrix}\,,
\end{align}
where $m_{1,2,3}$ are the masses of active neutrinos $\nu_{1,2,3}$.

%%%sec2b%%%
\subsection{Neutron-antineutron oscillation and proton decay}
	
\begin{figure}[t]
\centering
	\begin{tikzpicture}[line width=1pt, scale=1.5,>=Stealth]
	%%%%
	\path (0:0) coordinate (a0);
	\path (0:0.8) coordinate (a1);
	\path (180:0.8) coordinate (a2);
	\path (15:1.5) coordinate (a3);
	\path (345:1.42) coordinate (a4);
	\path (195:1.42) coordinate (a5);
	\path (165:1.5) coordinate (a6);
	\path (270:0.8) coordinate (a7);
	\path (330:1.7) coordinate (a8);
	\path (210:1.7) coordinate (a9);
	%%%%
	\draw (189:1.46) ellipse [x radius=0.06, y radius=0.62];
	\draw (351:1.46) ellipse [x radius=0.06, y radius=0.62];
	\draw[-](-1.8,0)--(-1.45,0);
	\draw[-](-1.8,-0.15)--(-1.45,-0.15);
	\draw[-](1.8,0)--(1.45,0);
	\draw[-](1.8,-0.15)--(1.45,-0.15);
	\draw [fermionbar](a1)--(a3);
	\draw [fermionbar](a1)--(a4);
	\draw [scalar](a1)--(a0);
	\draw [scalar](a2)--(a0);
	\draw [scalar](a7)--(a0);
	\draw [fermion](a6)--(a2);
	\draw [fermion](a5)--(a2);
	\draw [fermionbar](a7)--(a8);
	\draw [fermionbar](a7)--(a9);
	%%%%
	\node[left] at (165:1.5) {$u$};
	\node[left] at (195:1.5) {$d$};
	\node[right] at (15:1.5) {$\overline{u}$};
	\node[right] at (345:1.5) {$\overline{d}$};
	\node[above] at (0:0.5) {$S^{+1/3}$};
	\node[above] at (180:0.4) {$S^{+1/3}$};
	\node[right] at (270:0.4) {$\omega^{-2/3}$};
	\node[left] at (210:1.7) {$d$};
	\node[right] at (330:1.7) {$\overline{d}$};
	\node[right] at (357:1.8){$\overline{n}$};
	\node[left] at (183:1.8){$n$};
	\end{tikzpicture}\qquad
	\begin{tikzpicture}[line width=1pt, scale=1.5,>=Stealth]
	%%%%
	\path (0:0) coordinate (a0);
	\path (0:0.5) coordinate (a1);
	\path (180:0.5) coordinate (a2);
	\path (15:1.5) coordinate (a3);
	\path (345:1.5) coordinate (a4);
	\path (195:1.42) coordinate (a5);
	\path (165:1.5) coordinate (a6);
	\path (270:0.8) coordinate (a7);
	\path (330:1.7) coordinate (a8);
	\path (210:1.7) coordinate (a9);
	%%%%
	\draw (189:1.46) ellipse [x radius=0.06, y radius=0.62];
	\draw (337:1.56) ellipse [x radius=0.06, y radius=0.23];
	\draw[-](-1.8,0)--(-1.45,0);
	\draw[-](-1.8,-0.15)--(-1.45,-0.15);
	\draw[-](1.8,-0.55)--(1.45,-0.55);
	\draw[-](1.8,-0.7)--(1.45,-0.7);
	\draw [fermionbar](a1)--(a3);
	\draw [fermionbar](a1)--(a4);
	\draw [scalarbar](a1)--(a2);
	\draw [fermion](a6)--(a2);
	\draw [fermion](a5)--(a2);
	\draw [fermionbar](a8)--(a9);
	%%%%
	\node[left] at (165:1.5) {$u$};
	\node[left] at (195:1.55) {$d$};
	\node[right] at (15:1.5) {$e^{+}(\mu^{+})$};
	\node[right] at (345:1.5) {$\overline{u}$};
	\node[above] at (0:0) {$S^{+1/3}$};
	\node[left] at (210:1.7) {$u$};
	\node[right] at (330:1.7) {$u$};
	\node[right] at (342:1.9){$\pi^{0}$};
	\node[left] at (183:1.8){$p$};
	\end{tikzpicture}
\caption{Feynman diagrams of neutron-antineutron oscillation (left) and proton decay $p\to\pi^{0}e^{+}(\mu^{+})$ (right) in case 1 and case 2. For case 1, $S^{+1/3}=S_{1}^{1/3}$ and $\omega^{-2/3}=\omega_{1}^{-2/3}$. For case 2, $S^{+1/3}=S_{3}^{+1/3}$ and $\omega^{-2/3}=\omega_{3}^{-2/3}$.}\label{nnbar_pdecay}
\end{figure}
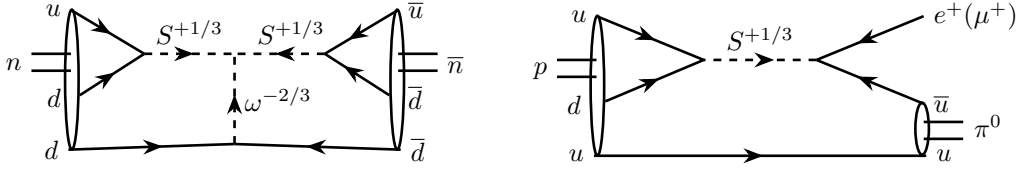

With the leptoquark coupling $z_{S}^{11}\neq0$ ($z_{S}$ corresponds to $z_{1SR}$ and $V^{*}z_{1SL}$ in case 1 and $V^{*}z_{3S}$ in case 2), case 1 and case 2 can lead to neutron-antineutron oscillation, as shown in Fig.~\ref{nnbar_pdecay}. The transition rate of neutron-antineutron oscillation $\tau^{-1}$ is proportional to $|(z_{S}^{11})^{2}z_{\omega}^{11}|$. Using the current limit $\tau\geq4.7\times10^{8}~{\rm{s}}$ given by the Super-Kamiokande (Super-K) experiment~\cite{Super-Kamiokande:2020bov}, one can get the bounds
\begin{align}
\left|(z_{S}^{11})^{2}z_{\omega}^{11}\right|\leq1.4\sim1.7\times 10^{-16}\times\left[\dfrac{M_{S}^{4}M^{2}_{\omega}}{\mu \cdot {\rm{TeV}}^{5}}\right]\,.
\end{align}
With $M_{S}\sim1.5~{\rm{TeV}}$ and $M_{\omega}\sim8~{\rm{TeV}}$, $|(z_{S}^{11})^{2}z_{\omega}^{11}|\lesssim 10^{-14}$. The construction of operators and detailed calculation of neutron-antineutron oscillation can be found in~\cite{Chang:1980ey,Kuo:1980ew,Rao:1982gt,Rao:1983sd,Caswell:1982qs,Buchoff:2015qwa,Rinaldi:2019thf,Oosterhof:2019dlo,Fridell:2021gag}. However, the proton will decay when $z_{S}^{11}$ and $y_{S}^{11}$ are both set nonzero. For example, the non-zero couplings can contribute to the process $p\rightarrow\pi^{0}e^{+}(\mu^{+})$ as shown in Fig.~\ref{nnbar_pdecay}. With the experimental limit $\tau/B(p\to \pi^{0}e^{+})>2.4\times10^{34}~{\rm{yrs}}$ and $\tau/B(p\to \pi^{0}\mu^{+})>1.6\times10^{34}~{\rm{yrs}}$ given by Super-K~\cite{Super-Kamiokande:2020wjk}, there is a strict bound on the couplings with the matrix element inputs from lattice~\cite{Aoki:2017puj,Yoo:2021gql}
\begin{align}
|z_{S}^{11}y_{S}^{11}|\lesssim10^{-26}\times\left[\dfrac{M_{S}}{\rm{TeV}}\right]^{2}\,.
\end{align}
where $y_{S}$ denotes $V^{*}y_{1SL}$ and $y_{1SR}$ in case 1 and $V^{*}y_{3S}$ in case 2. One can find that if $M_{S}\sim 1.5~{\rm{TeV}}$ and $y_{S}^{11}\sim0.01-1$, then $z_{S(L/R)}^{11}$ needs to be at $\mathcal{O}(10^{-26}-10^{-24})$ scale to avoid an inappropriate proton decay, leading to an unobservable neutron-antineutron oscillation. So in the discussion of cases 1 and 2, we assume $z_{S}=0$ to escape proton decay.

%%%sec2c%%%
	
\subsection{The texture setup and constraints}
We show our texture zeros setup of the couplings matrices and list the bounds on the couplings, which are related to $0\nu\beta\beta$ decay in this subsection. The bounds are derived from muon anomalous magnetic moment and tree level flavor violation processes with four-fermion interactions considered.
\subsubsection{Texture setup}
The standard parameterization of the PMNS matrix is
\begin{align}
U=\begin{pmatrix}
1&0&0\\
0&c_{23}&s_{23}\\
0&-s_{23}&c_{23}
\end{pmatrix}
\begin{pmatrix}
c_{13}&0&s_{13}e^{-i\delta}\\
0&1&0\\
-s_{13}e^{i\delta}&0&c_{13}
\end{pmatrix}
\begin{pmatrix}
c_{12}&s_{12}&0\\
-s_{12}&c_{12}&0\\
0&0&1		    
\end{pmatrix}
\begin{pmatrix}
1&0&0\\
0&e^{i\eta_{1}}&0\\
0&0&e^{i\eta_{2}}
\end{pmatrix}\,,
\end{align}
where $c_{ij}(s_{ij})$ denotes $\cos{\theta_{ij}}(\sin{\theta_{ij}})$, $\delta$ is the CP phase, and $\eta_{1,2}$ are the extra phases if neutrino are Majorana particles. The best fit values of these neutrino oscillation parameters have been derived in~\cite{Capozzi:2018ubv,Esteban:2020cvm,deSalas:2017kay, deSalas:2020pgw}. As there are no information about the Majorana phases ranges, they can varies from 0 to $2\pi$ freely. To evade constraints from various lepton flavor violation (LFV) processes, we adopt the Yukawa coupling matrices in case 1 as 
\begin{align}
y_{1SL}=V^{T}
\begin{pmatrix}
\textcolor{blue}{\#}&0&0\\
0&0&\#\\
\color{teal}{\#}&\textcolor{red}{\#}&\#\
\end{pmatrix}\,,\quad
y_{1SR}=\begin{pmatrix}
\textcolor{blue}{\#}&0&0\\
0&0&0\\
0&\textcolor{red}{\#}&0
\end{pmatrix}\,,\quad
z_{1\omega}=\begin{pmatrix}
\textcolor{blue}{\#}&0&0\\
0&0&\#\\
0&\#&\#
\end{pmatrix}\,.
\end{align}
The matrix $y_{1SL}$ and $z_{1\omega}$ are set to be complex and $y_{1SR}$ is real. The contribution to neutrinoless double beta decay can survive when couplings $(V^{*}y_{1SL})^{11}$ $\equiv y_{1SL}^{\prime 11}$, $y_{1SR}^{11}$ and $z_{\omega}^{11}$ (in blue) are set nonzero. Moreover, we have set $y_{1SL}^{\prime32}$, $y_{1SR}^{32}\neq 0$ (in red) to obtain the muon anomalous magnetic moment $(g-2)_{\mu}$. The entries of couplings $y_{1SL}^{\prime 23}, y_{1SL}^{\prime 33}, z_{1\omega}^{33}, z_{1\omega}^{23}$ provide enough independent parameters to generate appropriate neutrino mass matrix. It is noted that the first component of the neutrino mass matrix is negligible under these entries with the constraint from the tree-level flavor violation processes $|y_{1SL}^{\prime 11}|<0.12$, as shown in the following subsection, leading to an inconspicuous standard light neutrino exchange $0\nu\beta\beta$ decay. Hence we introduce $y_{1SL}^{\prime31}$ (in teal) to open the standard $0\nu\beta\beta$ decay.
 
The texture zeros setup of coupling matrices in case 2 and case 3 are similar to those in case 1
\begin{align}
y_{3S}=V^{T}\begin{pmatrix}\#&0&0\\0&0&\#\\
\#&\#&\#\end{pmatrix}\,,
z_{1\omega}=\begin{pmatrix}\#&0&0\\0&0&\#\\0&\#&\#\end{pmatrix}\,,
y_{2S}=\begin{pmatrix}\#&0&0\\0&0&\#\\
\#&\#&\#\end{pmatrix}\,,
z_{3\omega}=\begin{pmatrix}\#&0&0\\0&0&\#\\0&\#&\#\end{pmatrix}\,,
\end{align}
with $V^{*}y_{3S}\equiv y_{3S}^{\prime}$.
Though the coupling $|y_{2S}^{11}|$ could be $\mathcal{O}(1)$ in case 3, we still choose the same form of the matrix to make the analysis consistently.

\subsubsection{Constraints}
The Yukawa couplings are constrained by tree-level flavor violation processes. It is natural to work with effective field theory where the effective Lagrangian can be described with four-fermion interaction operators as the new particles are at TeV scale. The effective Lagrangian involving $S_{1}$ leptoquark reads~\cite{Chang:2016zll}
\begin{align}
\mathcal{L}_{{\rm{eff}},1}=&\dfrac{1}{2M^{2}_{S_{1}}}\bigg\{
(V^{*}y_{1SL})^{*ki}(V^{*}y_{1SL})^{nj}[\overline{E^{i}}\gamma_{\mu}P_{L}E^{j}][\overline{U^{k}}\gamma^{\mu}P_{L}U^{n}]\notag\\
&+y_{1SL}^{*ki}y_{1SL}^{nj}[\overline{\nu^{i}}\gamma_{\mu}P_{L}\nu^{j}][\overline{D^{k}}\gamma^{\mu}P_{L}D^{n}]
+y_{1SR}^{*ki}y_{1SR}^{nj}[\overline{E^{i}}\gamma_{\mu}P_{R}E^{j}][\overline{U^{k}}\gamma^{\mu}P_{R}U^{n}]\notag\\
&-\left[y_{1SL}^{*ki}(V^{*}y_{1SL})^{nj}[\overline{\nu^{i}}\gamma_{\mu}P_{L}E^{j}][\overline{D}^{k}\gamma^{\mu}P_{L}U^{n}]+{\rm{h.c.}}\right]\notag\\
&+\left[y_{1SR}^{*ki}y_{1SL}^{nj}[\overline{\nu^{i}}P_{R}E^{j}][\overline{D^{k}}P_{R}U^{n}]+{\rm{h.c.}}\right]\notag\\
&+\left[(V^{*}y_{1SL})^{*ki}y_{1SR}^{nj}[\overline{E^{i}}P_{R}E^{j}][\overline{U^{k}}P_{R}U^{n}]+{\rm{h.c.}}\right]\notag\\
&-\dfrac{1}{4}\left[y_{1SR}^{*ki}y_{1SL}^{nj}[\overline{\nu^{i}}\sigma_{\mu\nu}P_{R}E^{j}][\overline{D^{k}}\sigma^{\mu\nu}P_{R}U^{n}]+{\rm{h.c.}}\right]\notag\\
&-\dfrac{1}{4}\left[(V^{*}y_{1SL})^{*ki}y_{1SR}^{nj}[\overline{E^{i}}\sigma_{\mu\nu}P_{R}E^{j}][\overline{U^{k}}\sigma^{\mu\nu}P_{R}U^{n}]+{\rm{h.c.}}\right]\bigg\}\,.
\end{align}
The effective Lagrangians induced by $S_{3}$ and $\tilde{S}_{2}$ can be written as
\begin{align}
\mathcal{L}_{{\rm{eff}},2}=&\dfrac{1}{2M_{S_{3}}^{2}}\bigg\{
(V^{*}y_{3S})^{*ki}(V^{*}y_{3S})^{nj}[\overline{E^{i}}\gamma^{\mu}P_{L}E^{j}][\overline{U^{k}}\gamma_{\mu}P_{L}U^{n}]\notag\\
&+2(V^{*}y_{3S})^{*ki}(V^{*}y_{3S})^{nj}[\overline{\nu^{i}}\gamma^{\mu}P_{L}\nu^{j}][\overline{U^{k}}\gamma_{\mu}P_{L}U^{n}]\notag\\
&+2y_{3S}^{*ki}y_{3S}^{nj}[\overline{E^{i}}\gamma^{\mu}P_{L}E^{j}][\overline{D^{k}}\gamma_{\mu}P_{L}D^{n}]\notag\\
&+y_{3S}^{*ki}y_{3S}^{nj}[\overline{\nu^{i}}\gamma^{\mu}P_{L}\nu^{j}][\overline{D^{k}}\gamma_{\mu}P_{L}D^{n}]\notag\\
&+\left[y_{3S}^{*ki}(V^{*}y_{3S})^{nj}[\overline{\nu^{i}\gamma_{\mu}P_{L}E^{j}}][\overline{D^{k}}\gamma_{\mu}P_{L}U^{n}]+{\rm{h.c.}}\right]\bigg\}\,,\\
\mathcal{L}_{{\rm{eff}},3}=&-\dfrac{y_{2S}^{*ki}y_{2S}^{nj}}{2M_{S_{2}}^{2}}\bigg\{[\overline{E^{i}}\gamma^{\mu}P_{L}E^{j}][\overline{D^{k}}\gamma_{\mu}P_{R}D^{n}]
+[\overline{\nu^{i}}\gamma^{\mu}P_{L}\nu^{j}][\overline{D^{k}}\gamma_{\mu}P_{R}D^{n}]\bigg\}\,.
\end{align}
The constraints on the Wilson coefficient $\epsilon^{ijkn}$ have been derived in~\cite{Davidson:1993qk,Leurer:1993em,Leurer:1993qx,Carpentier:2010ue}, where
\begin{align}
|\epsilon^{ijkn}|={\rm{Const.}}\times\dfrac{\left|y_{S(L)}^{(\prime)ki}y_{S(L)}^{(\prime)nj}\right|}{4\sqrt{2}G_{F}M_{S}^{2}}\,,
\end{align}
with the constant equals $1$ in case 1 and case 3 while the constant takes as $1$ or $2$ in the case 2. Here we have taken the bounds from~\cite{Carpentier:2010ue} and have listed some of them in Table~\ref{constraint} which are related to the couplings that can contribute to neutrinoless double beta decay process. However, one needs to pay attention that bounds on the couplings $|y_{S(L)}^{\prime ki}y_{S(L)}^{\prime nj}|$ can be derived from the numerical combined calculation of the bounds on $|y_{S(L)}^{ki}y_{S(L)}^{nj}|$ and $|y_{S(L)}^{\prime ki}y_{S(L)}^{nj}|$ in case 1 and case 2. Moreover, the neutral meson mixing can be contributed by the diquarks. The $B^{0}_{d}-\bar{B}^{0}_{d}$ mixing needs to be concerned under our texture setup. The 95\% allowed range of the couplings~\cite{UTfit:2007eik} is
\begin{align}
|z_{1(3)\omega}^{11}z_{1(3)\omega}^{33}|<4.6(2.3)\times10^{-5}\times\left(\dfrac{M_{\omega1(3) }}{{\rm{TeV}}}\right)^{2}\,.
\end{align}
	
\begin{table}[t]
\centering
	\begin{tabular}{||c|c||c|c||c|c||}
	\hline
	Coefficients & Constraints & Coefficients & Constraints & Coefficients & Constraints \\
	\hline
	$|y_{1SL}^{\prime11}y_{1SL}^{\prime11}|$ & $6.88\times10^{-3}$ & $|y_{1SL}^{\prime11}y_{1SL}^{\prime31}|$ & $1.61\times10^{-2}$ & $|y_{1SL}^{\prime11}y_{1SL}^{\prime32}|$ & $2.57\times10^{-1}$ \\
	$|y_{1SL}^{\prime11}y_{1SL}^{\prime23}|$ & $3.10\times10^{-4}$ & $|y_{1SL}^{\prime11}y_{1SL}^{\prime33}|$ & $2.41\times10^{-1}$ & $|y_{1SR}^{11}y_{1SR}^{11}|$ & $6.5\times10^{-1}$ \\
	$|y_{1SR}^{11}y_{1SL}^{\prime32}|$ & $5.87$ & $|y_{1SR}^{11}y_{1SL}^{\prime23}|$ & $5.87$ &$|y_{1SR}^{11}y_{1SL}^{\prime33}|$ & $10.2$ \\
	\hline
	$|y_{3S}^{\prime11}y_{3S}^{\prime11}|$ & $6.78\times10^{-3}$ & $|y_{3S}^{\prime11}y_{3S}^{\prime31}|$ & $4.02\times10^{-3}$ & $|y_{3S}^{\prime11}y_{3S}^{\prime32}|$ & $4.66\times10^{-4}$ \\
	$|y_{3S}^{\prime11}y_{3S}^{\prime23}|$ & $3.10\times10^{-4}$ & $|y_{3S}^{\prime11}y_{3S}^{\prime33}|$ & $1.38\times10^{-1}$ & &\\
	\hline
	$|y_{2S}^{11}y_{2S}^{11}|$ & $1.78$ & $|y_{2S}^{11}y_{2S}^{31}|$ & $1.32\times10^{-2}$ & $|y_{2S}^{11}y_{2S}^{32}|$ & $1.32\times10^{-2}$\\
	$|y_{2S}^{11}y_{2S}^{23}|$ & $3.23\times10^{-1}$ & $|y_{2S}^{11}y_{2S}^{33}|$ & $2.71\times10^{-1}$ &&\\
	\hline
	\end{tabular}
\caption{The constraints on the couplings of the leptoquarks, $y_{1SL}^{\prime11},y_{1SR}^{11},y_{3SL}^{\prime11},y_{2S}^{11}$, which can contribute to $0\nu\beta\beta$ decay. The bounds are taken from~\cite{Carpentier:2010ue} with the unit of $(M_{S}/{\rm{TeV}})^{2}$.}\label{constraint}
\end{table}
The muon anomalous magnetic moments $a_{\mu}=(g-2)_{\mu}/2$ can also give information on the couplings. The latest result of muon anomalous magnetic moment has been presented by Muon $g-2$ collaboration~\cite{Muong-2:2021ojo} as
\begin{align}
\Delta a_{\mu}\equiv a_{\mu}^{\text{exp}}-a_{\mu}^{\text{SM}}=(2.51\pm0.59)\times 10^{-9}\,,
\end{align}
which has a 4.2 $\sigma$ discrepancy. The expression of muon anomalous magnetic moment in case 1 can be simplified as~\cite{Queiroz:2014zfa}
\begin{align}
\Delta a_{\mu}(S_{1})\simeq\dfrac{3m^{2}_{\mu}}{8\pi^{2}M_{S_{1}}^{2}}\dfrac{m_{t}}{m_{\mu}}{\rm{Re}}[y_{1SR}^{32}y_{ 1SL}^{\prime*32}]\left[\dfrac{1}{3}f_{1}\left(\dfrac{m_{t}^{2}}{m_{S_{1}}^{2}}\right)+\dfrac{2}{3}f_{2}\left(\dfrac{m_{t}^{2}}{m_{S_{1}}^{2}}\right)\right]\,.
\end{align}
While for case 2 and 3, the contributions are
\begin{align}
\Delta a_{\mu}(S_{3})\simeq\dfrac{3m^{2}_{\mu}}{8\pi^{2}M_{S_{3}}^{2}}\sum\limits_{q}|y_{3S}^{\prime q2}|^{2}\left[\dfrac{1}{3}f_{3}\left(\dfrac{m_{q}^{2}}{m_{S_{3}}^{2}}\right)+\dfrac{2}{3}f_{4}\left(\dfrac{m_{q}^{2}}{m_{S_{3}}^{2}}\right)\right]\,,\notag\\
\Delta a_{\mu}(\tilde{S}_{2})\simeq\dfrac{3m^{2}_{\mu}}{8\pi^{2}M_{S_{2}}^{2}}\sum\limits_{q}|y_{2S}^{q2}|^{2}\left[\dfrac{2}{3}f_{3}\left(\dfrac{m_{q}^{2}}{m_{S_{2}}^{2}}\right)+\dfrac{1}{3}f_{4}\left(\dfrac{m_{q}^{2}}{m_{S_{2}}^{2}}\right)\right]\,,
\end{align}
with the functions $f_{i}$ are defined in~\cite{Queiroz:2014zfa}. To explain the discrepancy, the couplings in case 1 have the relation ${\rm{Re}}[y_{1SR}^{32}y_{ 1SL}^{\prime*32}]\sim8.48\times10^{-2}$ with the leptoquark mass $M_{S_{1}}=1.5$ TeV. However, in case 2 and case 3, the contribution to muon anomalous magnetic moment $\Delta a_{\mu}$ are negligible since there is no chiral-enhancement $m_{q}/m_{\mu}$ for these two cases. 
	
After taking accounting of all the constraints mentioned, the parameter regions taken in our numerical analysis are
\begin{align}
{\rm{case~1:~}}& |y_{1SL}^{\prime11}|<0.12\,,~|y_{1SL}^{\prime31,32}|<0.3\,,~|y_{1SL}^{\prime33}|<0.4\,,~|y_{1SL}^{\prime23}|<0.005\,,\notag\\
&|y_{1SR}^{11}|<1.2\,,~{\rm{Re}}[y_{1SR}^{32}y_{1SL}^{\prime*32}]\sim0.0848\,,~|z_{1\omega}^{11}|<1.5\,,~|z_{1\omega}^{33}|<0.001\,,\\
{\rm{case~2:~}}&|y_{3S}^{\prime11}|<0.12\,,~|y_{3S}^{\prime31}|<0.07\,,~|y_{3S}^{\prime32}|<0.008\,,~|y_{3S}^{\prime33}|<0.4\,,\notag\\
&|y_{3S}^{\prime23}|<0.005\,,~|z_{1\omega}^{11}|<1.5\,,~|z_{1\omega}^{33}|<0.002\,,\\
{\rm{case~3:~}}&|y_{2S}^{11}|<1.5\,,~|y_{2S}^{31}|<0.01\,,~|y_{2S}^{32}|<0.01\,,~|y_{2S}^{33}|<0.3\,,\notag\\
&~|y_{2S}^{23}|<0.3\,,~|z_{3\omega}^{11}|<0.01\,,~|z_{3\omega}^{33}|<0.15\,,
\end{align}
and the $z_{i\omega}^{23}(i=1,3)$ in each case are set to be $|z_{i\omega}^{23}|<1.5$. We take $\mu=M_{S}=1.5~{\rm{TeV}}$ and $M_{\omega}=8$ TeV in our following discussion.

%%%sec3%%%
\section{The Neutrinoless Double Beta Decay}\label{sec3}
The $0\nu\beta\beta$ decay can be divided into short-range and long-range mechanisms. To study how short-range contributions impact the $0\nu\beta\beta$ decay in the cZB model, we briefly review the general formula of the short-range mechanisms via the effective field theory approach and give numerical analysis in this section.
	
%%%sec3A%%%
\subsection{The short-range $0\nu\beta\beta$ decay}
	
The short-range $0\nu\beta\beta$ decay operator can be written as $\mathcal{O}^{0\nu\beta\beta}\propto\overline{u}\overline{u}dd\overline{e}\overline{e}$, a dim-9 operator. The scalar mediated tree-level topologies and the decomposition of this operator has been listed in~\cite{Bonnet:2012kh}. We follow the general parameterization of effective short-range Lagrangian in~\cite{Deppisch:2020ztt,Pas:2000vn}
\begin{align}
\mathcal{L}_{SR}
=\dfrac{G_{F}^{2}V_{ud}^{2}}{2m_{p}}
\sum\limits_{X,Y,Z}&\bigg(\epsilon_{1}^{\chi}J_{X}J_{Y}j_{Z}
+\epsilon_{2}^{\chi}J^{\mu\nu}_{X}J_{Y,\mu\nu}j_{Z}
+\epsilon_{3}^{\chi}J^{\mu}_{X}J_{Y,\mu}j_{Z}\notag\\
&+\epsilon_{4}^{\chi}J^{\mu}_{X}J_{Y,\mu\nu}j^{\nu}
+\epsilon_{5}^{\chi}J^{\mu}_{X}J_{Y}j_{\mu}\bigg)+{\text{h.c.}}\\
=\dfrac{G_{F}^{2}V_{ud}^{2}}{2m_{p}}
\sum\limits_{\chi,i}&\epsilon_{i}^{\chi}\mathcal{O}_{i,\chi}^{0\nu\beta\beta}+{\text{h.c.}}\,, \label{L_SR} 
\end{align}    
where $G_{F}$ is the Fermi constant, $m_{p}$ is the proton mass, $V_{ud}$ is the $ud$ component of the CKM matrix, and the dimensionless effective couplings are defined as $\epsilon_{i}^{\chi}=\epsilon_{i}^{XYZ}~(X,Y,Z=R/L)$. The $J$ and $j$, respectively, denote the quark and electron currents as
\begin{equation}
\begin{gathered}
J_{R/L}=\overline{u}(1\pm\gamma_{5})d\,,\quad J_{R,L}^{\mu}=\overline{u}\gamma^{\mu}(1\pm\gamma_{5})d\,,\quad J_{R/L}^{\mu\nu}=\overline{u}\sigma^{\mu\nu}(1\pm\gamma_{5})d\,,\\
j_{R/L}=\overline{e}(1\mp\gamma_{5})e^{c},\quad j^{\mu}=\overline{e}\gamma^{\mu}\gamma_{5}e^{c}\,.
\end{gathered}
\end{equation}
One can express the effective operators in terms of the quark and electron currents as~\cite{Bonnet:2012kh,Graf:2018ozy}
\begin{equation}
\begin{gathered}
\mathcal{O}_{1,XYZ}^{0\nu\beta\beta}\equiv J_{X}J_{Y}j_{Z}\,,\quad
\mathcal{O}_{2,XYZ}^{0\nu\beta\beta}\equiv J^{\mu\nu}_{X}J_{Y,\mu\nu}j_{Z}\,,\quad
\mathcal{O}_{3,XYZ}^{0\nu\beta\beta}\equiv J^{\mu}_{X}J_{Y,\mu}j_{Z}\,,\\
\mathcal{O}_{4,XY}^{0\nu\beta\beta}\equiv J^{\mu}_{X}J_{Y,\mu\nu}j^{\nu}\,,\quad
\mathcal{O}_{5,XY}^{0\nu\beta\beta}\equiv J^{\mu}_{X}J_{Y}j_{\mu}\,.
\end{gathered}
\end{equation}
The following expression gives $0\nu\beta\beta$ decay inverse half-life involving the short-range mechanism and light-neutrino exchange~\cite{Deppisch:2020ztt}
\begin{align}
\left[T^{0\nu\beta\beta}_{1/2}\right]^{-1}
&=G_{11+}^{(0)}\left|\sum\limits_{i=1}^{3}\epsilon_{i}^{XYL}\mathcal{M}^{XY}_{i}+\epsilon_{\nu}\mathcal{M}_{\nu}\right|^{2}
+G_{11+}^{(0)}\left|\sum\limits_{i=1}^{3}\epsilon_{i}^{XYR}\mathcal{M}^{XY}_{i}\right|^{2}
+G_{66}^{(0)}\left|\sum\limits_{i=4}^{5}\epsilon_{i}^{XY}\mathcal{M}^{XY}_{i}\right|^{2}\notag\\
&+G_{16}^{(0)}\times 2{\text{Re}}\left[\left(\sum\limits_{i=1}^{3}\epsilon_{i}^{XYL}\mathcal{M}^{XY}_{i}-\sum\limits_{i=1}^{3}\epsilon_{i}^{XYR}\mathcal{M}^{XY}_{i}+\epsilon_{\nu}\mathcal{M}_{\nu}\right)\left(\sum\limits_{i=4}^{5}\epsilon_{i}^{XY}\mathcal{M}^{XY}_{i}\right)^{*}\right]\notag\\
&+G_{11-}^{(0)}\times2 {\text{Re}}\left[\left(\sum\limits_{i=1}^{3}\epsilon_{i}^{XYL}\mathcal{M}^{XY}_{i}+\epsilon_{\nu}\mathcal{M}_{\nu}\right)\left(\sum\limits_{i=1}^{3}\epsilon_{i}^{XYR}\mathcal{M}^{XY}_{i}\right)^{*}\right]\,,
\label{0vbb}
\end{align}
where $\epsilon_{i}^{\chi}$ are the effective couplings shown in Eq.~(\ref{L_SR}), $\mathcal{M}^{XY}_{i}$ are the NMEs with the short-range mechanism and $\mathcal{M}_{\nu}$ is the NMEs with the light neutrino exchange. The dimensionless parameters $\epsilon_{\nu}$ can be written as $\epsilon_{\nu}=\langle m_{ee}\rangle/m_{e}$, where $\langle m_{ee}\rangle\equiv\sum_{i}U_{ei}^{2}m_{i}$ is the effective Majorana neutrino mass and $m_{e}$ is the electron mass. The $G_{11+}^{(0)}$, $G_{11-}^{(0)}$, $G_{16}^{(0)}$, and $G_{66}^{(0)}$ are the phase space factors (PSFs) defined in~\cite{Kotila:2012zza}. The PSFs numerical values of different isotopes are taken from~\cite{Deppisch:2020ztt}, as shown in Table~\ref{PSF}. The first column is the lower limits for the decay half-life of different isotopes~\cite{GERDA:2020xhi,CUPID-0:2018rcs,CUPID:2020aow,Arnaboldi:2002te,CUORE:2021gpk,KamLAND-Zen:2016pfg}.  The Table~\ref{values-NMEs} shows the values of light neutrino exchange NME $\mathcal{M}_{\nu}$ and short-range mechanism NMEs $\mathcal{M}^{XY}_{i}$ within the microscopic interacting boson model~\cite{Deppisch:2020ztt}. We just list the values of $\mathcal{M}_{i}^{XX}$ since both of the quark currents in the cZB models are right-handed, i.e., $X=Y=R$. 

\begin{table}[t]
\centering
	\begin{tabular}{|c|c|cccc|}
	\hline
	Isotope&$T^{0\nu}_{1/2}$ ~[$10^{25}$ yrs] & $G_{11+}^{(0)}$ & $G_{11-}^{(0)}$ & $G_{16}^{(0)}$ & $G_{66}^{(0)}$\\
	\hline
	$^{76}$Ge&18~\cite{GERDA:2020xhi}&2.360&-0.280&0.870&1.320\\
	$^{82}$Se&0.24~\cite{CUPID-0:2018rcs}&10.19&-0.712&2.925&5.450\\
	$^{100}$Mo&0.15~\cite{CUPID:2020aow}&15.91&-1.053&4.456&8.482\\
	$^{128}$Te&0.011~\cite{Arnaboldi:2002te}&0.585&-0.156&0.313&0.371\\
	$^{130}$Te&2.2~\cite{CUORE:2021gpk}&14.20&-1.142&4.367&7.672\\
	$^{136}$Xe&10.7~\cite{KamLAND-Zen:2016pfg}&14.56&-1.197&4.524&7.876\\
	\hline
	\end{tabular}
\caption{The lower limits for the decay half-life time of different isotopes from different experiments and the numerical values of the phase space factors (PSFs) in units of $10^{-15}~{\text{yr}^{-1}}$~\cite{Deppisch:2020ztt}.}\label{PSF}
\end{table}
\begin{table}[t]
\centering
	\begin{tabular}{|c|cccccc|}
	\hline
	Isotope& $\mathcal{M}_{1}^{XX}$ & $\mathcal{M}_{2}^{XX}$ & $\mathcal{M}_{3}^{XX}$ & $\mathcal{M}_{4}^{XX}$ & $\mathcal{M}_{5}^{XX}$&$\mathcal{M}_{\nu}$\\
	\hline
	$^{76}$Ge&5300&-174&-200&-158&202&-6.64\\
	$^{82}$Se&4030&-144&-171&-134&114&-5.46\\
	$^{100}$Mo&12400&-189&-124&-134&1230&-5.27\\
	$^{128}$Te&4410&-134&-154&-130&205&-4.80\\
	$^{130}$Te&4030&-122&-141&-109&187&-4.40\\
	$^{136}$Xe&3210&-96.1&-111&-86.0&147&-3.60\\
	\hline
	\end{tabular}
\caption{The numerical values of the NMEs for short-range operators and light-neutrino exchange. The values are calculated within the microscopic interacting boson model and axial coupling quenched $g_{A}=1.0$~\cite{Deppisch:2020ztt}.}\label{values-NMEs}
\end{table}
	
\begin{figure}[!t]
\centering
	\begin{tikzpicture}[line width=0.8pt, scale=1.8,>=Stealth]
	%%%%
	\path (0:0) coordinate (a0);
	\path (150:1) coordinate (a1);
	\path (210:1) coordinate (a2);
	\path (0:0) coordinate (a3);
	\path (180:0.5) coordinate (a4);
	\path (30:1.12) coordinate (a5);
	\path (330:1.12) coordinate (a6);
	\path (30:0.5) coordinate (a7);
	\path (330:0.5) coordinate (a8);
	\path (15:1) coordinate (a9);
	\path (345:1) coordinate (a10);
	%%%%
	\draw [fermion](a1)--(a4);
	\draw [fermion](a2)--(a4);
	\draw [scalarbar](a3)--(a8);
	\draw [scalar](a4)--(a3);
	\draw [scalarbar](a3)--(a7);
	\draw [fermion](a7)--(a5);
	\draw [fermion](a7)--(a9);
	\draw [fermion](a8)--(a6);
	\draw [fermion](a8)--(a10);
	%%%%
	\node[left] at (150:1) {$d_{R}$};
	\node[left] at (210:1) {$d_{R}$};
	\node at (210:0.25) {$\omega_{1}$};
	\node at (70:0.3) {$S_{1}$};
	\node at (290:0.3) {$S_{1}$};
	\node[right] at (15:1) {$u_{L}$};
	\node[right] at (345:1) {$u_{L}$};
	\node[right] at (30:1.12) {$e_{L}$};
	\node[right] at (330:1.12) {$e_{L}$};
 	\end{tikzpicture}\quad
	\begin{tikzpicture}[line width=0.8pt, scale=1.8,>=Stealth]
	%%%%
	\path (0:0) coordinate (a0);
	\path (150:1) coordinate (a1);
	\path (210:1) coordinate (a2);
	\path (0:0) coordinate (a3);
	\path (180:0.5) coordinate (a4);
	\path (30:1.12) coordinate (a5);
	\path (330:1.12) coordinate (a6);
	\path (30:0.5) coordinate (a7);
	\path (330:0.5) coordinate (a8);
	\path (15:1) coordinate (a9);
	\path (345:1) coordinate (a10);
	%%%%
	\draw [fermion](a1)--(a4);
	\draw [fermion](a2)--(a4);
	\draw [scalarbar](a3)--(a8);
	\draw [scalar](a4)--(a3);
	\draw [scalarbar](a3)--(a7);
	\draw [fermion](a7)--(a5);
	\draw [fermion](a7)--(a9);
	\draw [fermion](a8)--(a6);
	\draw [fermion](a8)--(a10);
	%%%%
	\node[left] at (150:1) {$d_{R}$};
	\node[left] at (210:1) {$d_{R}$};
	\node at (210:0.25) {$\omega_{1}$};
	\node at (70:0.3) {$S_{1}$};
	\node at (290:0.3) {$S_{1}$};
	\node[right] at (15:1) {$u_{L}$};
	\node[right] at (345:1) {$u_{R}$};
	\node[right] at (30:1.12) {$e_{L}$};
	\node[right] at (330:1.12) {$e_{R}$};
	\end{tikzpicture}\quad
	\begin{tikzpicture}[line width=0.8pt, scale=1.8,>=Stealth]
	%%%%
	\path (0:0) coordinate (a0);
	\path (150:1) coordinate (a1);
	\path (210:1) coordinate (a2);
	\path (0:0) coordinate (a3);
	\path (180:0.5) coordinate (a4);
	\path (30:1.12) coordinate (a5);
	\path (330:1.12) coordinate (a6);
	\path (30:0.5) coordinate (a7);
	\path (330:0.5) coordinate (a8);
	\path (15:1) coordinate (a9);
	\path (345:1) coordinate (a10);
	%%%%
	\draw [fermion](a1)--(a4);
	\draw [fermion](a2)--(a4);
	\draw [scalarbar](a3)--(a8);
	\draw [scalar](a4)--(a3);
	\draw [scalarbar](a3)--(a7);
	\draw [fermion](a7)--(a5);
	\draw [fermion](a7)--(a9);
	\draw [fermion](a8)--(a6);
	\draw [fermion](a8)--(a10);
	%%%%
	\node[left] at (150:1) {$d_{R}$};
	\node[left] at (210:1) {$d_{R}$};
	\node at (210:0.25) {$\omega_{1}$};
	\node at (70:0.3) {$S_{1}$};
	\node at (290:0.3) {$S_{1}$};
	\node[right] at (15:1) {$u_{R}$};
	\node[right] at (345:1) {$u_{R}$};
	\node[right] at (30:1.12) {$e_{R}$};
	\node[right] at (330:1.12) {$e_{R}$};
	\end{tikzpicture}\\\vskip 0.4cm
	\begin{tikzpicture}[line width=0.8pt, scale=1.8,>=Stealth]
	%%%%
	\path (0:0) coordinate (a0);
	\path (150:1) coordinate (a1);
	\path (210:1) coordinate (a2);
	\path (0:0) coordinate (a3);
	\path (180:0.5) coordinate (a4);
	\path (30:1.12) coordinate (a5);
	\path (330:1.12) coordinate (a6);
	\path (30:0.5) coordinate (a7);
	\path (330:0.5) coordinate (a8);
	\path (15:1) coordinate (a9);
	\path (345:1) coordinate (a10);
	%%%%
	\draw [fermion](a1)--(a4);
	\draw [fermion](a2)--(a4);
	\draw [scalarbar](a3)--(a8);
	\draw [scalar](a4)--(a3);
	\draw [scalarbar](a3)--(a7);
	\draw [fermion](a7)--(a5);
	\draw [fermion](a7)--(a9);
	\draw [fermion](a8)--(a6);
	\draw [fermion](a8)--(a10);
	%%%%
	\node[left] at (150:1) {$d_{R}$};
	\node[left] at (210:1) {$d_{R}$};
	\node at (210:0.25) {$\omega_{1}$};
	\node at (70:0.3) {$S_{3}$};
	\node at (290:0.3) {$S_{3}$};
	\node[right] at (15:1) {$u_{L}$};
	\node[right] at (345:1) {$u_{L}$};
	\node[right] at (30:1.12) {$e_{L}$};
	\node[right] at (330:1.12) {$e_{L}$};
	\end{tikzpicture}\quad
	\begin{tikzpicture}[line width=0.8pt, scale=1.8,>=Stealth]
	%%%%
	\path (0:0) coordinate (a0);
	\path (90:0.5) coordinate (a1);
	\path (270:0.5) coordinate (a2);
	\path (150:1) coordinate (a3);
	\path (30:1.12) coordinate (a4);
	\path (330:1.12) coordinate (a5);
	\path (210:1) coordinate (a6);
	\path (0:0.5) coordinate (a7);
	\path (15:1) coordinate (a8);
	\path (345:1) coordinate (a9);
	%%%%
	\draw [scalarbar](a0)--(a1);
	\draw [scalarbar](a0)--(a2);
	\draw [scalar](a0)--(a7);
	\draw [fermion](a3)--(a1);
	\draw [fermion](a1)--(a4);
	\draw [fermion](a7)--(a9);
	\draw [fermion](a6)--(a2);
	\draw [fermion](a2)--(a5);
	\draw [fermion](a7)--(a8);
	%%%%
	\node[left] at (150:1) {$d_{R}$};
	\node[left] at (210:1) {$d_{R}$};
	\node at (30:0.25) {$\omega_{3}$};
	\node [left]at (90:0.3) {$\tilde{S}_{2}$};
	\node [left] at (270:0.3) {$\tilde{S}_{2}$};
	\node[right] at (15:1) {$u_{L}$};
	\node[right] at (345:1) {$u_{L}$};
	\node[right] at (30:1.12) {$e_{L}$};
	\node[right] at (330:1.12) {$e_{L}$};
	\end{tikzpicture}
\caption{The Feynman diagrams of neutrinoless double beta decay in the colored Zee-Babu model. The first row corresponds to case 1. The two diagrams in the second row coreespond to case 2 (left) and case 3 (right).}\label{NDBD}
\end{figure}
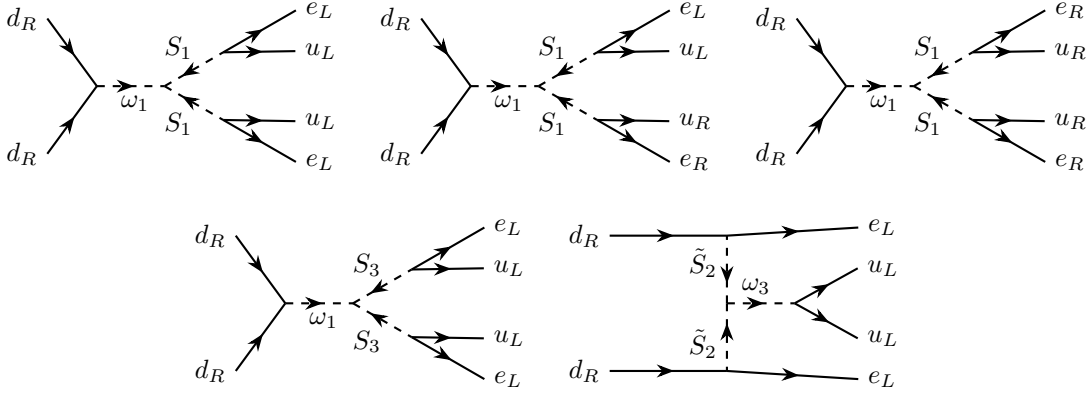

The effective operators in the three cases can be written as  
\begin{align}
{\rm{case~1:}}\quad&(\overline{u_{L}}\overline{e_{L}})(\overline{u_{L}}\overline{e_{L}})(d_{R}d_{R})\,
\rightarrow\dfrac{1}{48}\mathcal{O}_{1}^{RRL}-\dfrac{1}{192}\mathcal{O}_{2}^{RRL}\,,\notag\\ &(\overline{u_{L}}\overline{e_{L}})(\overline{u_{R}}\overline{e_{R}})(d_{R}d_{R})\,
\rightarrow\dfrac{1}{96i}\mathcal{O}_{4}^{RR}-\dfrac{1}{48}\mathcal{O}_{5}^{RR}\,,\notag\\ 
&(\overline{u_{R}}\overline{e_{R}})(\overline{u_{R}}\overline{e_{R}})(d_{R}d_{R})\,
\rightarrow-\dfrac{1}{48}\mathcal{O}_{3}^{RRR}\,,\\
{\rm{case~2:}}\quad&(\overline{u_{L}}\overline{e_{L}})(\overline{u_{L}}\overline{e_{L}})(d_{R}d_{R})\,
\rightarrow\dfrac{1}{48}\mathcal{O}_{1}^{RRL}-\dfrac{1}{192}\mathcal{O}_{2}^{RRL}\,,\\
{\rm{case~3:}}\quad&(\overline{u_{L}}\overline{u_{L}})(d_{R}\overline{e_{L}})(d_{R}\overline{e_{L}})\,
\rightarrow\dfrac{1}{48}\mathcal{O}_{1}^{RRL}-\dfrac{1}{192}\mathcal{O}_{2}^{RRL}\,.
\end{align}
The corresponding Feynman diagrams in different cases are shown in Fig.~\ref{NDBD}. 
The effective couplings $\epsilon_{i}^{\chi}$ in different cases are
\begin{align}
{\rm{case~1:}}\quad
&\epsilon_{1}^{RRL}
=+\dfrac{1}{48}\dfrac{2m_{p}}{G_{F}^{2}V_{ud}^{2}}\dfrac{4(y_{1SL}^{\prime*11})^{2}z_{1\omega}^{11}\mu_{1}}{M_{S_{1}}^{4}M^{2}_{\omega_{1}}}\,,\quad \epsilon_{2}^{RRL}=-\dfrac{1}{4}\epsilon_{1}^{RRL}\,,\nonumber\\
&\epsilon_{3}^{RRR}
=-\dfrac{1}{48}\dfrac{2m_{p}}{G_{F}^{2}V_{ud}^{2}}\dfrac{4(y_{1SR}^{*11})^{2}z_{1\omega}^{11}\mu_{1}}{M_{S_{1}}^{4}M^{2}_{\omega_{1}}}\,,
\nonumber\\
&\epsilon_{4}^{RR}
=+\dfrac{1}{96i}\dfrac{2m_{p}}{G_{F}^{2}V_{ud}^{2}}\dfrac{4y_{1SR}^{*11}y_{1SL}^{\prime*11}z_{1\omega}^{11}\mu_{1}}{M_{S_{1}}^{4}M^{2}_{\omega_{1}}}\,,\quad
\epsilon_{5}^{RR}=-2i\epsilon_{4}^{RR}\,,\\
{\rm{case~2:}}\quad
&\epsilon_{1}^{RRL}
=+\dfrac{1}{48}\dfrac{2m_{p}}{G_{F}^{2}V_{ud}^{2}}\dfrac{4(y_{3S}^{\prime*11})^{2}z_{1\omega}^{11}\mu_{2}}{M_{S_{3}}^{4}M^{2}_{\omega_{1}}}\,,\quad\epsilon_{2}^{RRL}=-\dfrac{1}{4}\epsilon_{1}^{RRL}\,,\\
{\rm{case~3:}}\quad
&\epsilon_{1}^{RRL}
=-\dfrac{1}{48}\dfrac{2m_{p}}{G_{F}^{2}V_{ud}^{2}}\dfrac{4(y_{2S}^{*11})^{2}(V^{*}z_{3\omega}V^{\dagger})^{11}\mu_{3}}{M_{S_{2}}^{4}M^{2}_{\omega_{3}}}\,,\quad\epsilon_{2}^{RRL}=-\dfrac{1}{4}\epsilon_{1}^{RRL}\,.
\end{align}
The effective neutrino mass can be related to the first component of the neutrino mass matrix~\cite{Helo:2015fba} and takes form as 
\begin{align}
|\langle m_{ee}\rangle|
&=\big|\sum\limits_{i}U_{ei}^{2}m_{i}\big|
=\big|M_{\nu}^{11}\big|\simeq\dfrac{3\mu m_{b}}{16\pi^{4}M_{S}^{2}}\tilde{I}\left(\dfrac{M_{\omega}^{2}}{M_{S}^{2}}\right)\left\{m_{s}|y_{S(L)}^{21}y_{S(L)}^{31}z_{\omega}^{23}|+m_{b}|[y_{S(L)}^{31}]^{2}z_{\omega}^{33}|\right\}\,.
\end{align}
The term related to $m_{s}$ cannot be neglected because we have assumed that there is a hierarchy among the couplings $z_{\omega}^{ij}$.

%%%sec3b%%%
\subsection{Numerical results}
Before we give our numerical results, it is necessary to notice that the QCD corrections can modify the NMEs~\cite{Mahajan:2013ixa,Gonzalez:2015ady,Arbelaez:2016zlt,Arbelaez:2016uto,Gonzalez:2017mcg,Ayala:2020gtv}. If we consider the leading order QCD corrections and the numerical values of the RGE $\mu$-evolution matrix elements with the same chiral quark currents~\cite{Gonzalez:2015ady}
\begin{align} 	
\hat{U}^{XX}_{(12)}=\begin{pmatrix}2.39&0.02\\-3.83&0.35\end{pmatrix}\,,\quad \hat{U}^{XX}_{(3)}=0.70\,,\quad
\hat{U}^{XX}_{(45)}=\begin{pmatrix}0.35&-0.96i\\-0.06i&2.39\end{pmatrix}\,,
\end{align}
the NMEs need to be recomposited as
\begin{align}
\mathcal{M}_{1}^{XX}\rightarrow\beta_{1}^{XX}&=2.39\mathcal{M}_{1}^{XX}-3.83\mathcal{M}_{2}^{XX}\,,\\
\mathcal{M}_{2}^{XX}\rightarrow\beta_{2}^{XX}&=0.02\mathcal{M}_{1}^{XX}+0.35\mathcal{M}_{2}^{XX}\,,\\
\mathcal{M}_{3}^{XX}\rightarrow\beta_{3}^{XX}&=0.70\mathcal{M}_{3}^{XX}\,,\\
\mathcal{M}_{4}^{XX}\rightarrow\beta_{4}^{XX}&=0.35\mathcal{M}_{4}^{XX}-0.06i\mathcal{M}_{5}^{XX}\,,\\
\mathcal{M}_{5}^{XX}\rightarrow\beta_{5}^{XX}&=-0.96i\mathcal{M}_{4}^{XX}+2.39\mathcal{M}_{5}^{XX}\,.
\end{align}
The inverse half-life $[T_{1/2}^{0\nu\beta\beta}]^{-1}(G_{jk},\epsilon_{i},\mathcal{M}_{i})$ have to be replaced with $[T_{1/2}^{0\nu\beta\beta}]^{-1}(G_{jk},\epsilon_{i},\beta_{i})$. After considering the experimental values and substituting the numerical values of the PSFs and the NMEs shown in Table~\ref{PSF} and \ref{values-NMEs}, one can get the limits on the couplings.
\begin{figure}[tb]
\centering
\includegraphics[height=9.5cm]{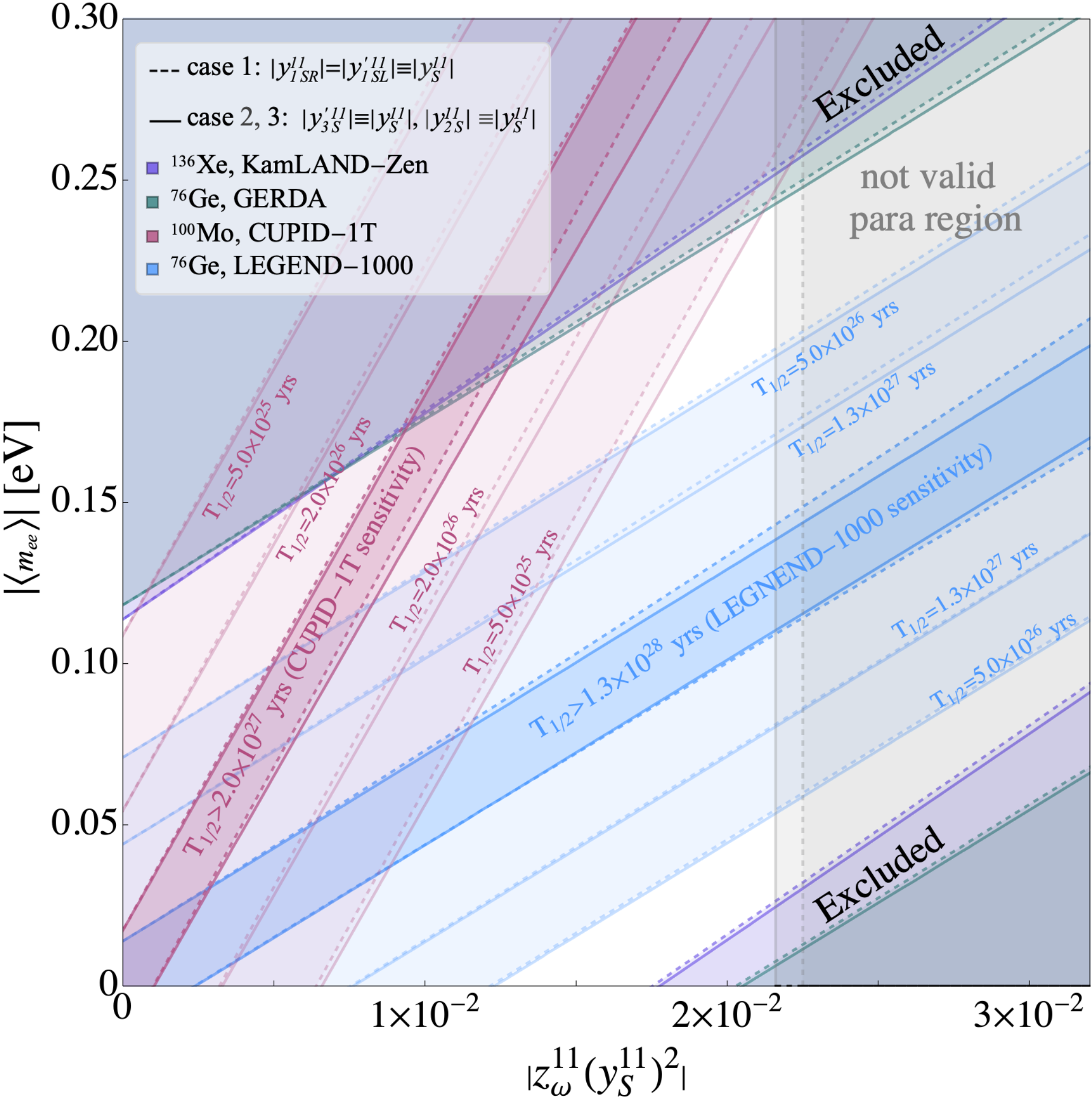}
\caption{The contour with the effective neutrino mass $|\langle m_{ee}\rangle|$ in the unit of ${\rm{eV}}$ and the couplings $|z_{\omega}^{11}(y_{S}^{11})^{2}|$. The lines show the experimental bound of $0\nu\beta\beta$ decay half-life in different cases and different isotopes. The upper and lower corner regions are excluded by KamLAND-Zen and GERDA experiments. The red and blue regions correspond to the experiments CUPID-1T and LEGEND-1000. The corresponding experiments are shown in the legends. The gray region is not valid in our numerical analysis. Here we take $\mu=M_{S}=1.5~{\rm{TeV}}$ and $M_{\omega}=8~{\rm{TeV}}$. We assume that $|y_{1SR}^{11}|=|y_{1SL}^{\prime 11}|$ in case 1 and denote $|y_{1SR}^{11}|$, $|y_{1SL}^{\prime 11}|$, $|y_{3S}^{\prime 11}|$, and $|y_{2S}^{11}|$ as $|y_{S}^{11}|$.}\label{plot-0vbb1}
\end{figure}
\begin{figure}[!htb]
\centering
\includegraphics[width=0.45\textwidth]{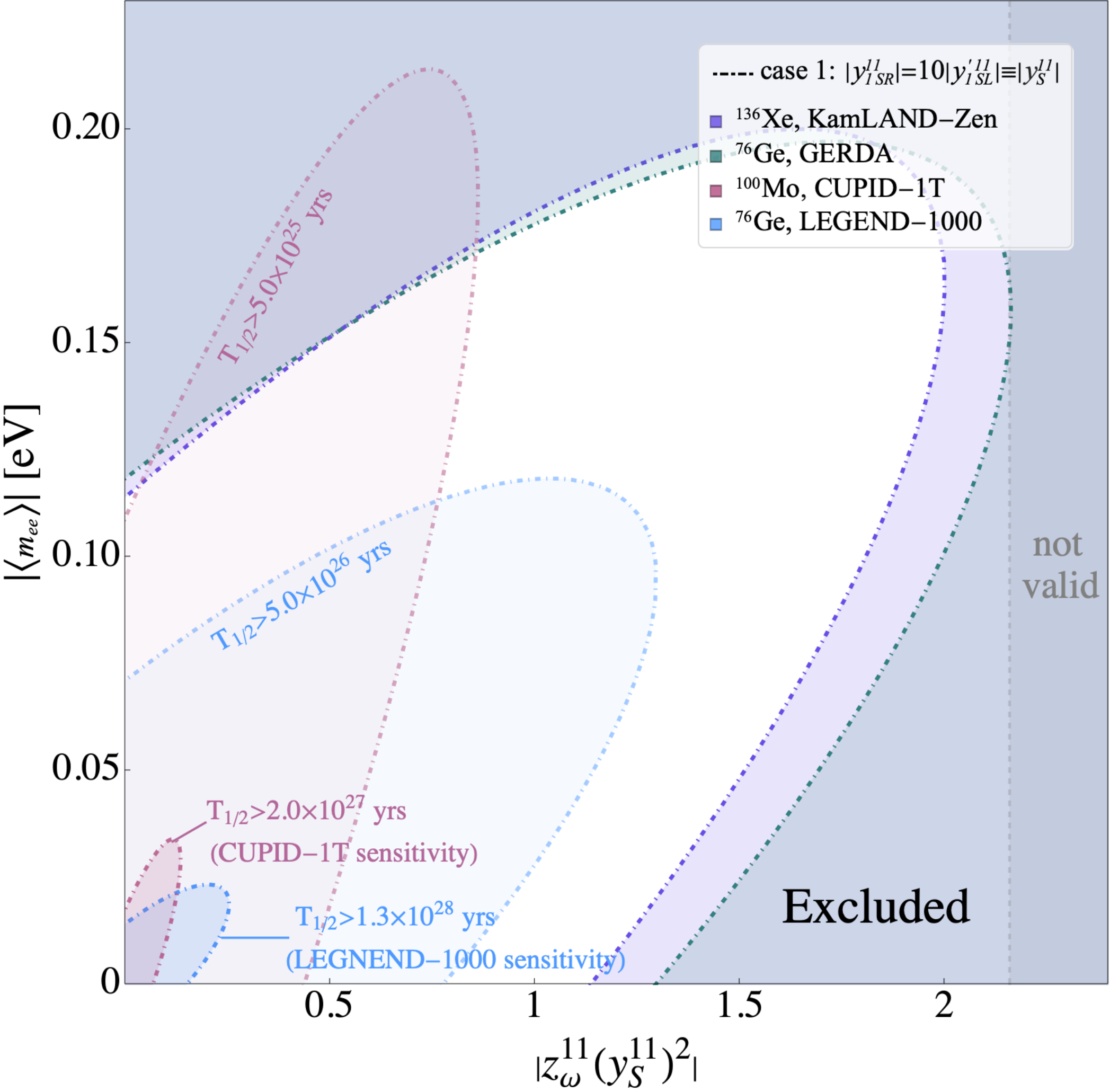}~~
\includegraphics[width=0.45\textwidth]{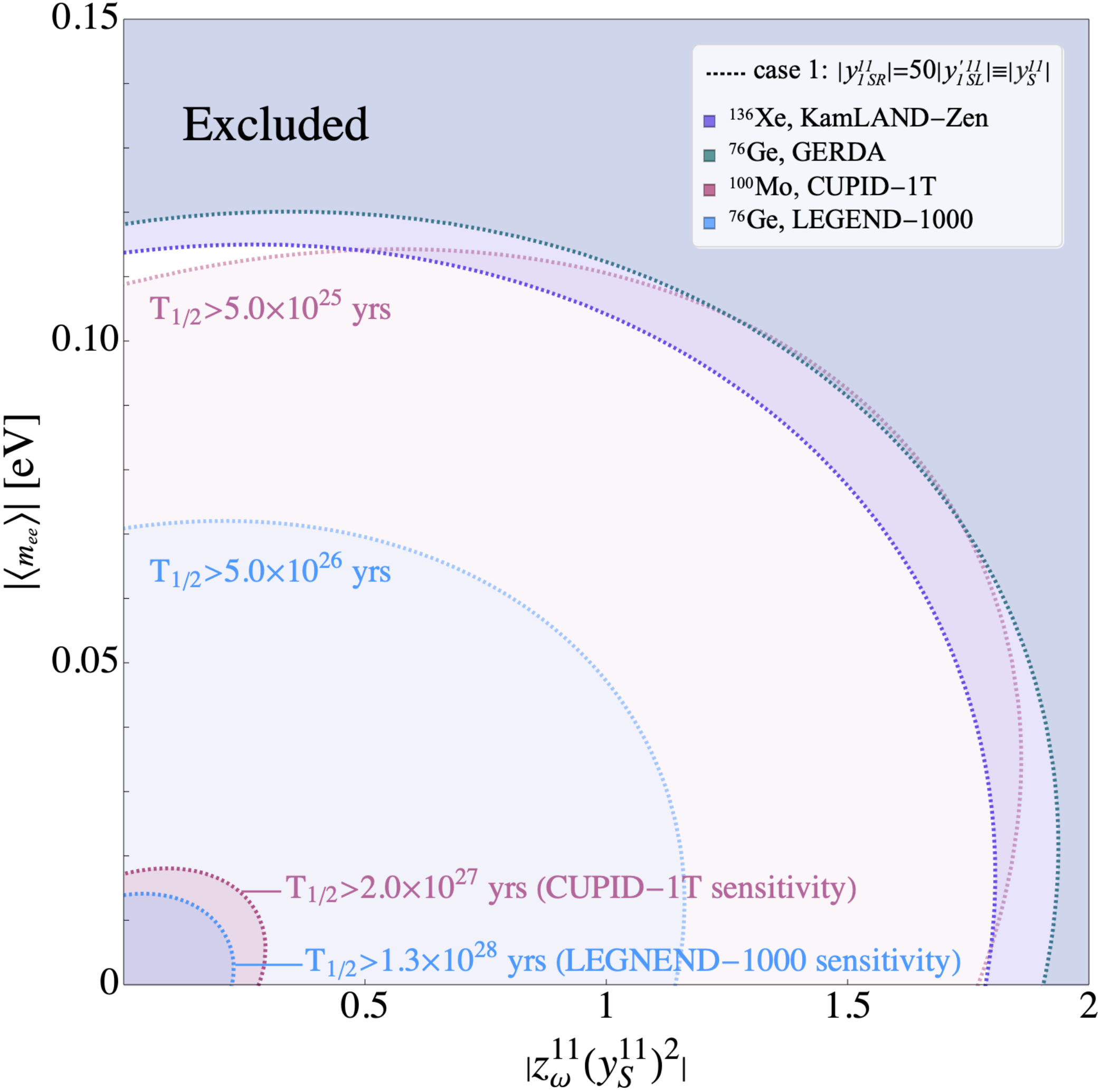}
\caption{The contour with the effective neutrino mass $|\langle m_{ee}\rangle|$ in the unit of ${\rm{eV}}$ and the couplings $|z_{\omega}^{11}(y_{S}^{11})^{2}|$ in case 1.
Here we take $\mu=M_{S}=1.5~{\rm{TeV}}$ and $M_{\omega}=8~{\rm{TeV}}$. We assume that $|y_{1SR}^{11}|=10|y_{1SL}^{\prime 11}|$(left) and $|y_{1SR}^{11}|=50|y_{1SL}^{\prime 11}|$(right), where $|y_{1SR}^{11}|$ is denoted as $|y_{S}^{11}|$.}\label{plot-0vbb2}
\end{figure}

We show the contour with the effective neutrino mass $|\langle m_{ee}\rangle|$ in the unit of ${\rm{eV}}$ and the couplings $|z_{\omega}^{11}(y_{S}^{11})^{2}|$ in Fig.~\ref{plot-0vbb1} and Fig.~\ref{plot-0vbb2}. The purple and green regions (upper and lower corners) are excluded by the $0\nu\beta\beta$ decay experiments KamLAND-Zen~\cite{KamLAND-Zen:2016pfg} and GERDA~\cite{GERDA:2020xhi}. The red and blue regions correspond to the survival areas for the experiments CUPID-1T~\cite{CUPID:2022wpt} and LEGEND-1000~\cite{LEGEND:2021bnm} if we assume that no signals are found and set the half-lifetime to $T_{1/2}>5.0\times 10^{25}$ yrs (CUPID) and $5.0\times 10^{26}$ yrs (LEGEND), whereas the inner darker areas relate to the sensitivities. We show all the three cases in Fig.~\ref{plot-0vbb1}, with the assumption $|y_{1SL}^{\prime11}|=|y_{1SR}^{11}|$ for case 1. Under this assumption, the term that contains $\epsilon_{1,2}^{RRL}$ and $\epsilon_{\nu}$ dominants in the inverse half-life expression. Due to the cancellation between the standard light neutrino exchange contribution and the new physics contribution, the $0\nu\beta\beta$ decay could be hidden when the relations are linear or almost linear, 
\begin{align}
(y_{S}^{*11})^{2}z_{\omega}^{11}\simeq300\times\dfrac{\mathcal{M}_{\nu}}{\beta_{1}-\beta_{2}/4}\dfrac{\langle m_{ee}\rangle}{\rm{eV}}\times\left(\dfrac{1.5~{\rm{TeV}}}{\mu}\right)\left(\dfrac{M_{S}}{1.5~{\rm{TeV}}}\right)^{4}\left(\dfrac{M_{\omega}}{8~{\rm{TeV}}}\right)^{2}\,,
\end{align}
where the specific value of $\mathcal{M}_{\nu}/(\beta_{1}-\beta_{2}/4)$ varies from isotope to isotope. For the current experiments, the relation can be realized successfully due to the similar ratio value of $\mathcal{M}_{\nu}/(\beta_{1}-\beta_{2}/4)$ in $^{76}{\rm{Ge}}$ and $^{136}{\rm{Xe}}$ isotopes. Things will be different and intriguing when the next-generation experiments with $^{100}{\rm{Mo}}$ isotope, e.g. AMoRE-II~\cite{Lee:2020rjh} and CUPID-1T, push the half-life to be order of $10^{26}~{\rm{yrs}}$. The slope of the band with $^{100}{\rm{Mo}}$ is different from $^{76}{\rm{Ge}}$ or $^{136}{\rm{Xe}}$, leading to the overlap of survival areas being narrowed. With the high sensitivity of the future CUPID-1T and LEGEND-1000 experiments, the survival band can be examined comprehensively. If there is no signal of $0\nu\beta\beta$ decays, the survival region will be reduced to the overlap area. On the other hand, if we see the signals in one experiment, the corresponding contour lines will be suitable. The other experiment can help us search for the appropriate region of the lines.
	
We show the contour of case 1 with assumption $|y_{1SL}^{\prime11}|\ll |y_{1SR}^{11}|\equiv |y_{S}^{11}|$ in Fig.~\ref{plot-0vbb2}. This assumption is natural as the allowed regions have ten times difference, and the influence on the neutrino mass is negligible. The survival region is elliptical instead. The left panel refers to $|y_{1SR}^{11}|=10|y_{1SL}^{\prime11}|$. One can find that the constraint on the effective Majorana neutrino mass can be larger than the one with only standard neutrino exchange considered $|\langle m_{ee}\rangle|\lesssim0.2~{\rm{eV}}$. The experiment with $^{100}{\rm{Mo}}$ isotopes can help to reduce the survival area which is similar to what we discussed before. The right panel refers to $|y_{1SR}^{11}|=50|y_{1SL}^{\prime11}|$ and the effect of the combined analysis with different experiments is not apparent. The bound on the couplings given by experiments KamLAND-Zen and GERDA are
\begin{align}
|z_{\omega}^{11}(y_{S}^{11})^{2}|
<1.8\times\left(\dfrac{1.5~{\text{TeV}}}{\mu}\right)
\left(\dfrac{M_{S}}{1.5~\text{TeV}}\right)^{4}
\left(\dfrac{M_{\omega}}{8~\text{TeV}}\right)^{2}\,.
\end{align}
The limits will be more stringent in next generation $0\nu\beta\beta$ decay experiments, which have the potential to restrict $|z_{\omega}^{11}(y_{S}^{11})^{2}|$ at $\mathcal{O}({10^{-1}})$ scale.

%%%sec4%%%
\section{Summary}\label{sec4}
In this paper, we have discussed the neutrinoless double beta decay in the colored Zee-Babu model. We study all three cases for the colored Zee-Babu model with a leptoquark and a diquark. The tiny neutrino masses are generated at two-loop level, and neutrinoless double beta decay gets additional contribution from the leptoquarks. We set some texture zeros for the Yukawa coupling matrices to evade constraints from various lepton flavor violation processes. We obtain the allowed regions of parameters after considering the constraints given by tree-level flavor violation processed and charged lepton anomalous magnetic moment.
	
We have discussed the short-range and standard neutrino exchange mechanisms of neutrinoless double beta decay for each case. The short-range contribution can be realized at tree-level. The general formula of the short-range contributions via the effective field theory approach is briefly reviewed. We adopt the values of nuclear matrix elements calculated with the microscopic interacting boson model and consider the leading order QCD running correction. We give numerical analysis for the three cases with the current experimental results and sensitivities of next-generation experiments. We find that the neutrinoless double beta decay can be hidden with a linear relation in all the cases under certain conditions. The relation can be examined by future $0\nu\beta\beta$ decay experiments. The complementary analysis of the different isotope experiments can help reduce the overlap area of the survival region.

%%%Acknowledgements%%%
\noindent{\bf Acknowledgements.}
This work is supported in part by the National Science Foundation of China (12175082, 11775093). 	

%%%references%%%
\bibliographystyle{apsrev}
\bibliography{references}

\begin{thebibliography}{85}
\expandafter\ifx\csname natexlab\endcsname\relax\def\natexlab#1{#1}\fi
\expandafter\ifx\csname bibnamefont\endcsname\relax
  \def\bibnamefont#1{#1}\fi
\expandafter\ifx\csname bibfnamefont\endcsname\relax
  \def\bibfnamefont#1{#1}\fi
\expandafter\ifx\csname citenamefont\endcsname\relax
  \def\citenamefont#1{#1}\fi
\expandafter\ifx\csname url\endcsname\relax
  \def\url#1{\texttt{#1}}\fi
\expandafter\ifx\csname urlprefix\endcsname\relax\def\urlprefix{URL }\fi
\providecommand{\bibinfo}[2]{#2}
\providecommand{\eprint}[2][]{\url{#2}}

\bibitem[{\citenamefont{Cheng and Li}(1980)}]{Cheng:1980qt}
\bibinfo{author}{\bibfnamefont{T.~P.} \bibnamefont{Cheng}} \bibnamefont{and}
  \bibinfo{author}{\bibfnamefont{L.-F.} \bibnamefont{Li}},
  \bibinfo{journal}{Phys. Rev. D} \textbf{\bibinfo{volume}{22}},
  \bibinfo{pages}{2860} (\bibinfo{year}{1980}).

\bibitem[{\citenamefont{Petcov and Toshev}(1984)}]{Petcov:1984nz}
\bibinfo{author}{\bibfnamefont{S.~T.} \bibnamefont{Petcov}} \bibnamefont{and}
  \bibinfo{author}{\bibfnamefont{S.~T.} \bibnamefont{Toshev}},
  \bibinfo{journal}{Phys. Lett. B} \textbf{\bibinfo{volume}{143}},
  \bibinfo{pages}{175} (\bibinfo{year}{1984}).

\bibitem[{\citenamefont{Zee}(1986)}]{Zee:1985id}
\bibinfo{author}{\bibfnamefont{A.}~\bibnamefont{Zee}}, \bibinfo{journal}{Nucl.
  Phys. B} \textbf{\bibinfo{volume}{264}}, \bibinfo{pages}{99}
  (\bibinfo{year}{1986}).

\bibitem[{\citenamefont{Babu}(1988)}]{Babu:1988ki}
\bibinfo{author}{\bibfnamefont{K.~S.} \bibnamefont{Babu}},
  \bibinfo{journal}{Phys. Lett. B} \textbf{\bibinfo{volume}{203}},
  \bibinfo{pages}{132} (\bibinfo{year}{1988}).

\bibitem[{\citenamefont{Babu and Julio}(2010)}]{Babu:2010vp}
\bibinfo{author}{\bibfnamefont{K.~S.} \bibnamefont{Babu}} \bibnamefont{and}
  \bibinfo{author}{\bibfnamefont{J.}~\bibnamefont{Julio}},
  \bibinfo{journal}{Nucl. Phys. B} \textbf{\bibinfo{volume}{841}},
  \bibinfo{pages}{130} (\bibinfo{year}{2010}), \eprint{1006.1092}.

\bibitem[{\citenamefont{Angel et~al.}(2013)\citenamefont{Angel, Cai, Rodd,
  Schmidt, and Volkas}}]{Angel:2013hla}
\bibinfo{author}{\bibfnamefont{P.~W.} \bibnamefont{Angel}},
  \bibinfo{author}{\bibfnamefont{Y.}~\bibnamefont{Cai}},
  \bibinfo{author}{\bibfnamefont{N.~L.} \bibnamefont{Rodd}},
  \bibinfo{author}{\bibfnamefont{M.~A.} \bibnamefont{Schmidt}},
  \bibnamefont{and} \bibinfo{author}{\bibfnamefont{R.~R.}
  \bibnamefont{Volkas}}, \bibinfo{journal}{JHEP} \textbf{\bibinfo{volume}{10}},
  \bibinfo{pages}{118} (\bibinfo{year}{2013}), \bibinfo{note}{[Erratum: JHEP
  11, 092 (2014)]}, \eprint{1308.0463}.

\bibitem[{\citenamefont{Aristizabal~Sierra
  et~al.}(2015)\citenamefont{Aristizabal~Sierra, Degee, Dorame, and
  Hirsch}}]{AristizabalSierra:2014wal}
\bibinfo{author}{\bibfnamefont{D.}~\bibnamefont{Aristizabal~Sierra}},
  \bibinfo{author}{\bibfnamefont{A.}~\bibnamefont{Degee}},
  \bibinfo{author}{\bibfnamefont{L.}~\bibnamefont{Dorame}}, \bibnamefont{and}
  \bibinfo{author}{\bibfnamefont{M.}~\bibnamefont{Hirsch}},
  \bibinfo{journal}{JHEP} \textbf{\bibinfo{volume}{03}}, \bibinfo{pages}{040}
  (\bibinfo{year}{2015}), \eprint{1411.7038}.

\bibitem[{\citenamefont{Cao et~al.}(2018)\citenamefont{Cao, Chen, Ma, Yan, and
  Zhang}}]{Cao:2017xgk}
\bibinfo{author}{\bibfnamefont{Q.-H.} \bibnamefont{Cao}},
  \bibinfo{author}{\bibfnamefont{S.-L.} \bibnamefont{Chen}},
  \bibinfo{author}{\bibfnamefont{E.}~\bibnamefont{Ma}},
  \bibinfo{author}{\bibfnamefont{B.}~\bibnamefont{Yan}}, \bibnamefont{and}
  \bibinfo{author}{\bibfnamefont{D.-M.} \bibnamefont{Zhang}},
  \bibinfo{journal}{Phys. Lett. B} \textbf{\bibinfo{volume}{779}},
  \bibinfo{pages}{430} (\bibinfo{year}{2018}), \eprint{1707.05896}.

\bibitem[{\citenamefont{Gando et~al.}(2016)}]{KamLAND-Zen:2016pfg}
\bibinfo{author}{\bibfnamefont{A.}~\bibnamefont{Gando}} \bibnamefont{et~al.}
  (\bibinfo{collaboration}{KamLAND-Zen}), \bibinfo{journal}{Phys. Rev. Lett.}
  \textbf{\bibinfo{volume}{117}}, \bibinfo{pages}{082503}
  (\bibinfo{year}{2016}), \bibinfo{note}{[Addendum: Phys.Rev.Lett. 117, 109903
  (2016)]}, \eprint{1605.02889}.

\bibitem[{\citenamefont{Agostini et~al.}(2020)}]{GERDA:2020xhi}
\bibinfo{author}{\bibfnamefont{M.}~\bibnamefont{Agostini}} \bibnamefont{et~al.}
  (\bibinfo{collaboration}{GERDA}), \bibinfo{journal}{Phys. Rev. Lett.}
  \textbf{\bibinfo{volume}{125}}, \bibinfo{pages}{252502}
  (\bibinfo{year}{2020}), \eprint{2009.06079}.

\bibitem[{\citenamefont{Armatol et~al.}(2022)}]{CUPID:2022wpt}
\bibinfo{author}{\bibfnamefont{A.}~\bibnamefont{Armatol}} \bibnamefont{et~al.}
  (\bibinfo{collaboration}{CUPID}) (\bibinfo{year}{2022}), \eprint{2203.08386}.

\bibitem[{\citenamefont{Abgrall et~al.}(2021)}]{LEGEND:2021bnm}
\bibinfo{author}{\bibfnamefont{N.}~\bibnamefont{Abgrall}} \bibnamefont{et~al.}
  (\bibinfo{collaboration}{LEGEND}) (\bibinfo{year}{2021}),
  \eprint{2107.11462}.

\bibitem[{\citenamefont{Prezeau et~al.}(2003)\citenamefont{Prezeau,
  Ramsey-Musolf, and Vogel}}]{Prezeau:2003xn}
\bibinfo{author}{\bibfnamefont{G.}~\bibnamefont{Prezeau}},
  \bibinfo{author}{\bibfnamefont{M.}~\bibnamefont{Ramsey-Musolf}},
  \bibnamefont{and} \bibinfo{author}{\bibfnamefont{P.}~\bibnamefont{Vogel}},
  \bibinfo{journal}{Phys. Rev. D} \textbf{\bibinfo{volume}{68}},
  \bibinfo{pages}{034016} (\bibinfo{year}{2003}), \eprint{hep-ph/0303205}.

\bibitem[{\citenamefont{del Aguila et~al.}(2012)\citenamefont{del Aguila,
  Aparici, Bhattacharya, Santamaria, and Wudka}}]{delAguila:2012nu}
\bibinfo{author}{\bibfnamefont{F.}~\bibnamefont{del Aguila}},
  \bibinfo{author}{\bibfnamefont{A.}~\bibnamefont{Aparici}},
  \bibinfo{author}{\bibfnamefont{S.}~\bibnamefont{Bhattacharya}},
  \bibinfo{author}{\bibfnamefont{A.}~\bibnamefont{Santamaria}},
  \bibnamefont{and} \bibinfo{author}{\bibfnamefont{J.}~\bibnamefont{Wudka}},
  \bibinfo{journal}{JHEP} \textbf{\bibinfo{volume}{06}}, \bibinfo{pages}{146}
  (\bibinfo{year}{2012}), \eprint{1204.5986}.

\bibitem[{\citenamefont{Cirigliano et~al.}(2017)\citenamefont{Cirigliano,
  Dekens, de~Vries, Graesser, and Mereghetti}}]{Cirigliano:2017djv}
\bibinfo{author}{\bibfnamefont{V.}~\bibnamefont{Cirigliano}},
  \bibinfo{author}{\bibfnamefont{W.}~\bibnamefont{Dekens}},
  \bibinfo{author}{\bibfnamefont{J.}~\bibnamefont{de~Vries}},
  \bibinfo{author}{\bibfnamefont{M.~L.} \bibnamefont{Graesser}},
  \bibnamefont{and}
  \bibinfo{author}{\bibfnamefont{E.}~\bibnamefont{Mereghetti}},
  \bibinfo{journal}{JHEP} \textbf{\bibinfo{volume}{12}}, \bibinfo{pages}{082}
  (\bibinfo{year}{2017}), \eprint{1708.09390}.

\bibitem[{\citenamefont{Deppisch et~al.}(2018)\citenamefont{Deppisch, Graf,
  Harz, and Huang}}]{Deppisch:2017ecm}
\bibinfo{author}{\bibfnamefont{F.~F.} \bibnamefont{Deppisch}},
  \bibinfo{author}{\bibfnamefont{L.}~\bibnamefont{Graf}},
  \bibinfo{author}{\bibfnamefont{J.}~\bibnamefont{Harz}}, \bibnamefont{and}
  \bibinfo{author}{\bibfnamefont{W.-C.} \bibnamefont{Huang}},
  \bibinfo{journal}{Phys. Rev. D} \textbf{\bibinfo{volume}{98}},
  \bibinfo{pages}{055029} (\bibinfo{year}{2018}), \eprint{1711.10432}.

\bibitem[{\citenamefont{Cirigliano et~al.}(2018)\citenamefont{Cirigliano,
  Dekens, de~Vries, Graesser, and Mereghetti}}]{Cirigliano:2018yza}
\bibinfo{author}{\bibfnamefont{V.}~\bibnamefont{Cirigliano}},
  \bibinfo{author}{\bibfnamefont{W.}~\bibnamefont{Dekens}},
  \bibinfo{author}{\bibfnamefont{J.}~\bibnamefont{de~Vries}},
  \bibinfo{author}{\bibfnamefont{M.~L.} \bibnamefont{Graesser}},
  \bibnamefont{and}
  \bibinfo{author}{\bibfnamefont{E.}~\bibnamefont{Mereghetti}},
  \bibinfo{journal}{JHEP} \textbf{\bibinfo{volume}{12}}, \bibinfo{pages}{097}
  (\bibinfo{year}{2018}), \eprint{1806.02780}.

\bibitem[{\citenamefont{Gr\'af et~al.}(2022)\citenamefont{Gr\'af, Lindner, and
  Scholer}}]{Graf:2022lhj}
\bibinfo{author}{\bibfnamefont{L.}~\bibnamefont{Gr\'af}},
  \bibinfo{author}{\bibfnamefont{M.}~\bibnamefont{Lindner}}, \bibnamefont{and}
  \bibinfo{author}{\bibfnamefont{O.}~\bibnamefont{Scholer}}
  (\bibinfo{year}{2022}), \eprint{2204.10845}.

\bibitem[{\citenamefont{Pas et~al.}(1998)\citenamefont{Pas, Hirsch, Kovalenko,
  and Klapdor-Kleingrothaus}}]{Pas:1997cp}
\bibinfo{author}{\bibfnamefont{H.}~\bibnamefont{Pas}},
  \bibinfo{author}{\bibfnamefont{M.}~\bibnamefont{Hirsch}},
  \bibinfo{author}{\bibfnamefont{S.~G.} \bibnamefont{Kovalenko}},
  \bibnamefont{and} \bibinfo{author}{\bibfnamefont{H.~V.}
  \bibnamefont{Klapdor-Kleingrothaus}}, \bibinfo{journal}{Prog. Part. Nucl.
  Phys.} \textbf{\bibinfo{volume}{40}}, \bibinfo{pages}{283}
  (\bibinfo{year}{1998}), \eprint{hep-ph/9712361}.

\bibitem[{\citenamefont{Pas et~al.}(1999)\citenamefont{Pas, Hirsch,
  Klapdor-Kleingrothaus, and Kovalenko}}]{Pas:1999fc}
\bibinfo{author}{\bibfnamefont{H.}~\bibnamefont{Pas}},
  \bibinfo{author}{\bibfnamefont{M.}~\bibnamefont{Hirsch}},
  \bibinfo{author}{\bibfnamefont{H.~V.} \bibnamefont{Klapdor-Kleingrothaus}},
  \bibnamefont{and} \bibinfo{author}{\bibfnamefont{S.~G.}
  \bibnamefont{Kovalenko}}, \bibinfo{journal}{Phys. Lett. B}
  \textbf{\bibinfo{volume}{453}}, \bibinfo{pages}{194} (\bibinfo{year}{1999}).

\bibitem[{\citenamefont{Helo et~al.}(2016)\citenamefont{Helo, Hirsch, and
  Ota}}]{Helo:2016vsi}
\bibinfo{author}{\bibfnamefont{J.~C.} \bibnamefont{Helo}},
  \bibinfo{author}{\bibfnamefont{M.}~\bibnamefont{Hirsch}}, \bibnamefont{and}
  \bibinfo{author}{\bibfnamefont{T.}~\bibnamefont{Ota}},
  \bibinfo{journal}{JHEP} \textbf{\bibinfo{volume}{06}}, \bibinfo{pages}{006}
  (\bibinfo{year}{2016}), \eprint{1602.03362}.

\bibitem[{\citenamefont{Kotila et~al.}(2021)\citenamefont{Kotila, Ferretti, and
  Iachello}}]{Kotila:2021xgw}
\bibinfo{author}{\bibfnamefont{J.}~\bibnamefont{Kotila}},
  \bibinfo{author}{\bibfnamefont{J.}~\bibnamefont{Ferretti}}, \bibnamefont{and}
  \bibinfo{author}{\bibfnamefont{F.}~\bibnamefont{Iachello}}
  (\bibinfo{year}{2021}), \eprint{2110.09141}.

\bibitem[{\citenamefont{Pas et~al.}(2001)\citenamefont{Pas, Hirsch,
  Klapdor-Kleingrothaus, and Kovalenko}}]{Pas:2000vn}
\bibinfo{author}{\bibfnamefont{H.}~\bibnamefont{Pas}},
  \bibinfo{author}{\bibfnamefont{M.}~\bibnamefont{Hirsch}},
  \bibinfo{author}{\bibfnamefont{H.~V.} \bibnamefont{Klapdor-Kleingrothaus}},
  \bibnamefont{and} \bibinfo{author}{\bibfnamefont{S.~G.}
  \bibnamefont{Kovalenko}}, \bibinfo{journal}{Phys. Lett. B}
  \textbf{\bibinfo{volume}{498}}, \bibinfo{pages}{35} (\bibinfo{year}{2001}),
  \eprint{hep-ph/0008182}.

\bibitem[{\citenamefont{Graf et~al.}(2018)\citenamefont{Graf, Deppisch,
  Iachello, and Kotila}}]{Graf:2018ozy}
\bibinfo{author}{\bibfnamefont{L.}~\bibnamefont{Graf}},
  \bibinfo{author}{\bibfnamefont{F.~F.} \bibnamefont{Deppisch}},
  \bibinfo{author}{\bibfnamefont{F.}~\bibnamefont{Iachello}}, \bibnamefont{and}
  \bibinfo{author}{\bibfnamefont{J.}~\bibnamefont{Kotila}},
  \bibinfo{journal}{Phys. Rev. D} \textbf{\bibinfo{volume}{98}},
  \bibinfo{pages}{095023} (\bibinfo{year}{2018}), \eprint{1806.06058}.

\bibitem[{\citenamefont{Bonnet et~al.}(2013)\citenamefont{Bonnet, Hirsch, Ota,
  and Winter}}]{Bonnet:2012kh}
\bibinfo{author}{\bibfnamefont{F.}~\bibnamefont{Bonnet}},
  \bibinfo{author}{\bibfnamefont{M.}~\bibnamefont{Hirsch}},
  \bibinfo{author}{\bibfnamefont{T.}~\bibnamefont{Ota}}, \bibnamefont{and}
  \bibinfo{author}{\bibfnamefont{W.}~\bibnamefont{Winter}},
  \bibinfo{journal}{JHEP} \textbf{\bibinfo{volume}{03}}, \bibinfo{pages}{055}
  (\bibinfo{year}{2013}), \bibinfo{note}{[Erratum: JHEP 04, 090 (2014)]},
  \eprint{1212.3045}.

\bibitem[{\citenamefont{Chen et~al.}(2021)\citenamefont{Chen, Ding, and
  Yao}}]{Chen:2021rcv}
\bibinfo{author}{\bibfnamefont{P.-T.} \bibnamefont{Chen}},
  \bibinfo{author}{\bibfnamefont{G.-J.} \bibnamefont{Ding}}, \bibnamefont{and}
  \bibinfo{author}{\bibfnamefont{C.-Y.} \bibnamefont{Yao}},
  \bibinfo{journal}{JHEP} \textbf{\bibinfo{volume}{12}}, \bibinfo{pages}{169}
  (\bibinfo{year}{2021}), \eprint{2110.15347}.

\bibitem[{\citenamefont{Helo et~al.}(2013)\citenamefont{Helo, Hirsch, P\"as,
  and Kovalenko}}]{Helo:2013ika}
\bibinfo{author}{\bibfnamefont{J.~C.} \bibnamefont{Helo}},
  \bibinfo{author}{\bibfnamefont{M.}~\bibnamefont{Hirsch}},
  \bibinfo{author}{\bibfnamefont{H.}~\bibnamefont{P\"as}}, \bibnamefont{and}
  \bibinfo{author}{\bibfnamefont{S.~G.} \bibnamefont{Kovalenko}},
  \bibinfo{journal}{Phys. Rev. D} \textbf{\bibinfo{volume}{88}},
  \bibinfo{pages}{073011} (\bibinfo{year}{2013}), \eprint{1307.4849}.

\bibitem[{\citenamefont{Deppisch et~al.}(2020)\citenamefont{Deppisch, Graf,
  Iachello, and Kotila}}]{Deppisch:2020ztt}
\bibinfo{author}{\bibfnamefont{F.~F.} \bibnamefont{Deppisch}},
  \bibinfo{author}{\bibfnamefont{L.}~\bibnamefont{Graf}},
  \bibinfo{author}{\bibfnamefont{F.}~\bibnamefont{Iachello}}, \bibnamefont{and}
  \bibinfo{author}{\bibfnamefont{J.}~\bibnamefont{Kotila}},
  \bibinfo{journal}{Phys. Rev. D} \textbf{\bibinfo{volume}{102}},
  \bibinfo{pages}{095016} (\bibinfo{year}{2020}), \eprint{2009.10119}.

\bibitem[{\citenamefont{Kohda et~al.}(2013)\citenamefont{Kohda, Sugiyama, and
  Tsumura}}]{Kohda:2012sr}
\bibinfo{author}{\bibfnamefont{M.}~\bibnamefont{Kohda}},
  \bibinfo{author}{\bibfnamefont{H.}~\bibnamefont{Sugiyama}}, \bibnamefont{and}
  \bibinfo{author}{\bibfnamefont{K.}~\bibnamefont{Tsumura}},
  \bibinfo{journal}{Phys. Lett. B} \textbf{\bibinfo{volume}{718}},
  \bibinfo{pages}{1436} (\bibinfo{year}{2013}), \eprint{1210.5622}.

\bibitem[{\citenamefont{Nomura and Okada}(2016)}]{Nomura:2016ask}
\bibinfo{author}{\bibfnamefont{T.}~\bibnamefont{Nomura}} \bibnamefont{and}
  \bibinfo{author}{\bibfnamefont{H.}~\bibnamefont{Okada}},
  \bibinfo{journal}{Phys. Rev. D} \textbf{\bibinfo{volume}{94}},
  \bibinfo{pages}{075021} (\bibinfo{year}{2016}), \eprint{1607.04952}.

\bibitem[{\citenamefont{Chang et~al.}(2016)\citenamefont{Chang, Liou, Wong, and
  Xu}}]{Chang:2016zll}
\bibinfo{author}{\bibfnamefont{W.-F.} \bibnamefont{Chang}},
  \bibinfo{author}{\bibfnamefont{S.-C.} \bibnamefont{Liou}},
  \bibinfo{author}{\bibfnamefont{C.-F.} \bibnamefont{Wong}}, \bibnamefont{and}
  \bibinfo{author}{\bibfnamefont{F.}~\bibnamefont{Xu}}, \bibinfo{journal}{JHEP}
  \textbf{\bibinfo{volume}{10}}, \bibinfo{pages}{106} (\bibinfo{year}{2016}),
  \eprint{1608.05511}.

\bibitem[{\citenamefont{Guo et~al.}(2018)\citenamefont{Guo, Han, Li, Liao, and
  Ma}}]{Guo:2017gxp}
\bibinfo{author}{\bibfnamefont{S.-Y.} \bibnamefont{Guo}},
  \bibinfo{author}{\bibfnamefont{Z.-L.} \bibnamefont{Han}},
  \bibinfo{author}{\bibfnamefont{B.}~\bibnamefont{Li}},
  \bibinfo{author}{\bibfnamefont{Y.}~\bibnamefont{Liao}}, \bibnamefont{and}
  \bibinfo{author}{\bibfnamefont{X.-D.} \bibnamefont{Ma}},
  \bibinfo{journal}{Nucl. Phys. B} \textbf{\bibinfo{volume}{928}},
  \bibinfo{pages}{435} (\bibinfo{year}{2018}), \eprint{1707.00522}.

\bibitem[{\citenamefont{Ding et~al.}(2018)\citenamefont{Ding, Han, Huang, and
  Liao}}]{Ding:2018jdk}
\bibinfo{author}{\bibfnamefont{R.}~\bibnamefont{Ding}},
  \bibinfo{author}{\bibfnamefont{Z.-L.} \bibnamefont{Han}},
  \bibinfo{author}{\bibfnamefont{L.}~\bibnamefont{Huang}}, \bibnamefont{and}
  \bibinfo{author}{\bibfnamefont{Y.}~\bibnamefont{Liao}},
  \bibinfo{journal}{Chin. Phys. C} \textbf{\bibinfo{volume}{42}},
  \bibinfo{pages}{103101} (\bibinfo{year}{2018}), \eprint{1802.05248}.

\bibitem[{\citenamefont{Datta et~al.}(2019)\citenamefont{Datta, Sachdeva, and
  Waite}}]{Datta:2019tuj}
\bibinfo{author}{\bibfnamefont{A.}~\bibnamefont{Datta}},
  \bibinfo{author}{\bibfnamefont{D.}~\bibnamefont{Sachdeva}}, \bibnamefont{and}
  \bibinfo{author}{\bibfnamefont{J.}~\bibnamefont{Waite}},
  \bibinfo{journal}{Phys. Rev. D} \textbf{\bibinfo{volume}{100}},
  \bibinfo{pages}{055015} (\bibinfo{year}{2019}), \eprint{1905.04046}.

\bibitem[{\citenamefont{Saad}(2020)}]{Saad:2020ihm}
\bibinfo{author}{\bibfnamefont{S.}~\bibnamefont{Saad}}, \bibinfo{journal}{Phys.
  Rev. D} \textbf{\bibinfo{volume}{102}}, \bibinfo{pages}{015019}
  (\bibinfo{year}{2020}), \eprint{2005.04352}.

\bibitem[{\citenamefont{Babu et~al.}(2021)\citenamefont{Babu, Dev, Jana, and
  Thapa}}]{Babu:2020hun}
\bibinfo{author}{\bibfnamefont{K.~S.} \bibnamefont{Babu}},
  \bibinfo{author}{\bibfnamefont{P.~S.~B.} \bibnamefont{Dev}},
  \bibinfo{author}{\bibfnamefont{S.}~\bibnamefont{Jana}}, \bibnamefont{and}
  \bibinfo{author}{\bibfnamefont{A.}~\bibnamefont{Thapa}},
  \bibinfo{journal}{JHEP} \textbf{\bibinfo{volume}{03}}, \bibinfo{pages}{179}
  (\bibinfo{year}{2021}), \eprint{2009.01771}.

\bibitem[{\citenamefont{Hirsch et~al.}(1996{\natexlab{a}})\citenamefont{Hirsch,
  Klapdor-Kleingrothaus, and Kovalenko}}]{Hirsch:1996qy}
\bibinfo{author}{\bibfnamefont{M.}~\bibnamefont{Hirsch}},
  \bibinfo{author}{\bibfnamefont{H.~V.} \bibnamefont{Klapdor-Kleingrothaus}},
  \bibnamefont{and} \bibinfo{author}{\bibfnamefont{S.~G.}
  \bibnamefont{Kovalenko}}, \bibinfo{journal}{Phys. Lett. B}
  \textbf{\bibinfo{volume}{378}}, \bibinfo{pages}{17}
  (\bibinfo{year}{1996}{\natexlab{a}}), \eprint{hep-ph/9602305}.

\bibitem[{\citenamefont{Hirsch et~al.}(1996{\natexlab{b}})\citenamefont{Hirsch,
  Klapdor-Kleingrothaus, and Kovalenko}}]{Hirsch:1996ye}
\bibinfo{author}{\bibfnamefont{M.}~\bibnamefont{Hirsch}},
  \bibinfo{author}{\bibfnamefont{H.~V.} \bibnamefont{Klapdor-Kleingrothaus}},
  \bibnamefont{and} \bibinfo{author}{\bibfnamefont{S.~G.}
  \bibnamefont{Kovalenko}}, \bibinfo{journal}{Phys. Rev. D}
  \textbf{\bibinfo{volume}{54}}, \bibinfo{pages}{R4207}
  (\bibinfo{year}{1996}{\natexlab{b}}), \eprint{hep-ph/9603213}.

\bibitem[{\citenamefont{Davies and He}(1991)}]{Davies:1990sc}
\bibinfo{author}{\bibfnamefont{A.~J.} \bibnamefont{Davies}} \bibnamefont{and}
  \bibinfo{author}{\bibfnamefont{X.-G.} \bibnamefont{He}},
  \bibinfo{journal}{Phys. Rev. D} \textbf{\bibinfo{volume}{43}},
  \bibinfo{pages}{225} (\bibinfo{year}{1991}).

\bibitem[{\citenamefont{Dor\v{s}ner et~al.}(2016)\citenamefont{Dor\v{s}ner,
  Fajfer, Greljo, Kamenik, and Ko\v{s}nik}}]{Dorsner:2016wpm}
\bibinfo{author}{\bibfnamefont{I.}~\bibnamefont{Dor\v{s}ner}},
  \bibinfo{author}{\bibfnamefont{S.}~\bibnamefont{Fajfer}},
  \bibinfo{author}{\bibfnamefont{A.}~\bibnamefont{Greljo}},
  \bibinfo{author}{\bibfnamefont{J.~F.} \bibnamefont{Kamenik}},
  \bibnamefont{and}
  \bibinfo{author}{\bibfnamefont{N.}~\bibnamefont{Ko\v{s}nik}},
  \bibinfo{journal}{Phys. Rept.} \textbf{\bibinfo{volume}{641}},
  \bibinfo{pages}{1} (\bibinfo{year}{2016}), \eprint{1603.04993}.

\bibitem[{\citenamefont{Sirunyan et~al.}(2019{\natexlab{a}})}]{CMS:2018ncu}
\bibinfo{author}{\bibfnamefont{A.~M.} \bibnamefont{Sirunyan}}
  \bibnamefont{et~al.} (\bibinfo{collaboration}{CMS}), \bibinfo{journal}{Phys.
  Rev. D} \textbf{\bibinfo{volume}{99}}, \bibinfo{pages}{052002}
  (\bibinfo{year}{2019}{\natexlab{a}}), \eprint{1811.01197}.

\bibitem[{\citenamefont{Aad et~al.}(2020)}]{ATLAS:2020dsk}
\bibinfo{author}{\bibfnamefont{G.}~\bibnamefont{Aad}} \bibnamefont{et~al.}
  (\bibinfo{collaboration}{ATLAS}), \bibinfo{journal}{JHEP}
  \textbf{\bibinfo{volume}{10}}, \bibinfo{pages}{112} (\bibinfo{year}{2020}),
  \eprint{2006.05872}.

\bibitem[{\citenamefont{Aaboud et~al.}(2019)}]{ATLAS:2019ebv}
\bibinfo{author}{\bibfnamefont{M.}~\bibnamefont{Aaboud}} \bibnamefont{et~al.}
  (\bibinfo{collaboration}{ATLAS}), \bibinfo{journal}{Eur. Phys. J. C}
  \textbf{\bibinfo{volume}{79}}, \bibinfo{pages}{733} (\bibinfo{year}{2019}),
  \eprint{1902.00377}.

\bibitem[{\citenamefont{Sirunyan et~al.}(2019{\natexlab{b}})}]{CMS:2018lab}
\bibinfo{author}{\bibfnamefont{A.~M.} \bibnamefont{Sirunyan}}
  \bibnamefont{et~al.} (\bibinfo{collaboration}{CMS}), \bibinfo{journal}{Phys.
  Rev. D} \textbf{\bibinfo{volume}{99}}, \bibinfo{pages}{032014}
  (\bibinfo{year}{2019}{\natexlab{b}}), \eprint{1808.05082}.

\bibitem[{\citenamefont{Sirunyan et~al.}(2018)}]{CMS:2018svy}
\bibinfo{author}{\bibfnamefont{A.~M.} \bibnamefont{Sirunyan}}
  \bibnamefont{et~al.} (\bibinfo{collaboration}{CMS}), \bibinfo{journal}{Eur.
  Phys. J. C} \textbf{\bibinfo{volume}{78}}, \bibinfo{pages}{707}
  (\bibinfo{year}{2018}), \eprint{1803.02864}.

\bibitem[{\citenamefont{Chatrchyan et~al.}(2012)}]{CMS:2012cyn}
\bibinfo{author}{\bibfnamefont{S.}~\bibnamefont{Chatrchyan}}
  \bibnamefont{et~al.} (\bibinfo{collaboration}{CMS}), \bibinfo{journal}{JHEP}
  \textbf{\bibinfo{volume}{12}}, \bibinfo{pages}{055} (\bibinfo{year}{2012}),
  \eprint{1210.5627}.

\bibitem[{\citenamefont{Sirunyan et~al.}(2020)}]{CMS:2019gwf}
\bibinfo{author}{\bibfnamefont{A.~M.} \bibnamefont{Sirunyan}}
  \bibnamefont{et~al.} (\bibinfo{collaboration}{CMS}), \bibinfo{journal}{JHEP}
  \textbf{\bibinfo{volume}{05}}, \bibinfo{pages}{033} (\bibinfo{year}{2020}),
  \eprint{1911.03947}.

\bibitem[{\citenamefont{Babu and Macesanu}(2003)}]{Babu:2002uu}
\bibinfo{author}{\bibfnamefont{K.~S.} \bibnamefont{Babu}} \bibnamefont{and}
  \bibinfo{author}{\bibfnamefont{C.}~\bibnamefont{Macesanu}},
  \bibinfo{journal}{Phys. Rev. D} \textbf{\bibinfo{volume}{67}},
  \bibinfo{pages}{073010} (\bibinfo{year}{2003}), \eprint{hep-ph/0212058}.

\bibitem[{\citenamefont{Abe et~al.}(2021)}]{Super-Kamiokande:2020bov}
\bibinfo{author}{\bibfnamefont{K.}~\bibnamefont{Abe}} \bibnamefont{et~al.}
  (\bibinfo{collaboration}{Super-Kamiokande}), \bibinfo{journal}{Phys. Rev. D}
  \textbf{\bibinfo{volume}{103}}, \bibinfo{pages}{012008}
  (\bibinfo{year}{2021}), \eprint{2012.02607}.

\bibitem[{\citenamefont{Chang and Chang}(1980)}]{Chang:1980ey}
\bibinfo{author}{\bibfnamefont{L.~N.} \bibnamefont{Chang}} \bibnamefont{and}
  \bibinfo{author}{\bibfnamefont{N.~P.} \bibnamefont{Chang}},
  \bibinfo{journal}{Phys. Lett. B} \textbf{\bibinfo{volume}{92}},
  \bibinfo{pages}{103} (\bibinfo{year}{1980}).

\bibitem[{\citenamefont{Kuo and Love}(1980)}]{Kuo:1980ew}
\bibinfo{author}{\bibfnamefont{T.-K.} \bibnamefont{Kuo}} \bibnamefont{and}
  \bibinfo{author}{\bibfnamefont{S.~T.} \bibnamefont{Love}},
  \bibinfo{journal}{Phys. Rev. Lett.} \textbf{\bibinfo{volume}{45}},
  \bibinfo{pages}{93} (\bibinfo{year}{1980}).

\bibitem[{\citenamefont{Rao and Shrock}(1982)}]{Rao:1982gt}
\bibinfo{author}{\bibfnamefont{S.}~\bibnamefont{Rao}} \bibnamefont{and}
  \bibinfo{author}{\bibfnamefont{R.}~\bibnamefont{Shrock}},
  \bibinfo{journal}{Phys. Lett. B} \textbf{\bibinfo{volume}{116}},
  \bibinfo{pages}{238} (\bibinfo{year}{1982}).

\bibitem[{\citenamefont{Rao and Shrock}(1984)}]{Rao:1983sd}
\bibinfo{author}{\bibfnamefont{S.}~\bibnamefont{Rao}} \bibnamefont{and}
  \bibinfo{author}{\bibfnamefont{R.~E.} \bibnamefont{Shrock}},
  \bibinfo{journal}{Nucl. Phys. B} \textbf{\bibinfo{volume}{232}},
  \bibinfo{pages}{143} (\bibinfo{year}{1984}).

\bibitem[{\citenamefont{Caswell et~al.}(1983)\citenamefont{Caswell,
  Milutinovic, and Senjanovic}}]{Caswell:1982qs}
\bibinfo{author}{\bibfnamefont{W.~E.} \bibnamefont{Caswell}},
  \bibinfo{author}{\bibfnamefont{J.}~\bibnamefont{Milutinovic}},
  \bibnamefont{and}
  \bibinfo{author}{\bibfnamefont{G.}~\bibnamefont{Senjanovic}},
  \bibinfo{journal}{Phys. Lett. B} \textbf{\bibinfo{volume}{122}},
  \bibinfo{pages}{373} (\bibinfo{year}{1983}).

\bibitem[{\citenamefont{Buchoff and Wagman}(2016)}]{Buchoff:2015qwa}
\bibinfo{author}{\bibfnamefont{M.~I.} \bibnamefont{Buchoff}} \bibnamefont{and}
  \bibinfo{author}{\bibfnamefont{M.}~\bibnamefont{Wagman}},
  \bibinfo{journal}{Phys. Rev. D} \textbf{\bibinfo{volume}{93}},
  \bibinfo{pages}{016005} (\bibinfo{year}{2016}), \bibinfo{note}{[Erratum:
  Phys.Rev.D 98, 079901 (2018)]}, \eprint{1506.00647}.

\bibitem[{\citenamefont{Rinaldi et~al.}(2019)\citenamefont{Rinaldi, Syritsyn,
  Wagman, Buchoff, Schroeder, and Wasem}}]{Rinaldi:2019thf}
\bibinfo{author}{\bibfnamefont{E.}~\bibnamefont{Rinaldi}},
  \bibinfo{author}{\bibfnamefont{S.}~\bibnamefont{Syritsyn}},
  \bibinfo{author}{\bibfnamefont{M.~L.} \bibnamefont{Wagman}},
  \bibinfo{author}{\bibfnamefont{M.~I.} \bibnamefont{Buchoff}},
  \bibinfo{author}{\bibfnamefont{C.}~\bibnamefont{Schroeder}},
  \bibnamefont{and} \bibinfo{author}{\bibfnamefont{J.}~\bibnamefont{Wasem}},
  \bibinfo{journal}{Phys. Rev. D} \textbf{\bibinfo{volume}{99}},
  \bibinfo{pages}{074510} (\bibinfo{year}{2019}), \eprint{1901.07519}.

\bibitem[{\citenamefont{Oosterhof et~al.}(2019)\citenamefont{Oosterhof, Long,
  de~Vries, Timmermans, and van Kolck}}]{Oosterhof:2019dlo}
\bibinfo{author}{\bibfnamefont{F.}~\bibnamefont{Oosterhof}},
  \bibinfo{author}{\bibfnamefont{B.}~\bibnamefont{Long}},
  \bibinfo{author}{\bibfnamefont{J.}~\bibnamefont{de~Vries}},
  \bibinfo{author}{\bibfnamefont{R.~G.~E.} \bibnamefont{Timmermans}},
  \bibnamefont{and} \bibinfo{author}{\bibfnamefont{U.}~\bibnamefont{van
  Kolck}}, \bibinfo{journal}{Phys. Rev. Lett.} \textbf{\bibinfo{volume}{122}},
  \bibinfo{pages}{172501} (\bibinfo{year}{2019}), \eprint{1902.05342}.

\bibitem[{\citenamefont{Fridell et~al.}(2021)\citenamefont{Fridell, Harz, and
  Hati}}]{Fridell:2021gag}
\bibinfo{author}{\bibfnamefont{K.}~\bibnamefont{Fridell}},
  \bibinfo{author}{\bibfnamefont{J.}~\bibnamefont{Harz}}, \bibnamefont{and}
  \bibinfo{author}{\bibfnamefont{C.}~\bibnamefont{Hati}},
  \bibinfo{journal}{JHEP} \textbf{\bibinfo{volume}{11}}, \bibinfo{pages}{185}
  (\bibinfo{year}{2021}), \eprint{2105.06487}.

\bibitem[{\citenamefont{Takenaka et~al.}(2020)}]{Super-Kamiokande:2020wjk}
\bibinfo{author}{\bibfnamefont{A.}~\bibnamefont{Takenaka}} \bibnamefont{et~al.}
  (\bibinfo{collaboration}{Super-Kamiokande}), \bibinfo{journal}{Phys. Rev. D}
  \textbf{\bibinfo{volume}{102}}, \bibinfo{pages}{112011}
  (\bibinfo{year}{2020}), \eprint{2010.16098}.

\bibitem[{\citenamefont{Aoki et~al.}(2017)\citenamefont{Aoki, Izubuchi,
  Shintani, and Soni}}]{Aoki:2017puj}
\bibinfo{author}{\bibfnamefont{Y.}~\bibnamefont{Aoki}},
  \bibinfo{author}{\bibfnamefont{T.}~\bibnamefont{Izubuchi}},
  \bibinfo{author}{\bibfnamefont{E.}~\bibnamefont{Shintani}}, \bibnamefont{and}
  \bibinfo{author}{\bibfnamefont{A.}~\bibnamefont{Soni}},
  \bibinfo{journal}{Phys. Rev. D} \textbf{\bibinfo{volume}{96}},
  \bibinfo{pages}{014506} (\bibinfo{year}{2017}), \eprint{1705.01338}.

\bibitem[{\citenamefont{Yoo et~al.}(2022)\citenamefont{Yoo, Aoki, Boyle,
  Izubuchi, Soni, and Syritsyn}}]{Yoo:2021gql}
\bibinfo{author}{\bibfnamefont{J.-S.} \bibnamefont{Yoo}},
  \bibinfo{author}{\bibfnamefont{Y.}~\bibnamefont{Aoki}},
  \bibinfo{author}{\bibfnamefont{P.}~\bibnamefont{Boyle}},
  \bibinfo{author}{\bibfnamefont{T.}~\bibnamefont{Izubuchi}},
  \bibinfo{author}{\bibfnamefont{A.}~\bibnamefont{Soni}}, \bibnamefont{and}
  \bibinfo{author}{\bibfnamefont{S.}~\bibnamefont{Syritsyn}},
  \bibinfo{journal}{Phys. Rev. D} \textbf{\bibinfo{volume}{105}},
  \bibinfo{pages}{074501} (\bibinfo{year}{2022}), \eprint{2111.01608}.

\bibitem[{\citenamefont{Capozzi et~al.}(2018)\citenamefont{Capozzi, Lisi,
  Marrone, and Palazzo}}]{Capozzi:2018ubv}
\bibinfo{author}{\bibfnamefont{F.}~\bibnamefont{Capozzi}},
  \bibinfo{author}{\bibfnamefont{E.}~\bibnamefont{Lisi}},
  \bibinfo{author}{\bibfnamefont{A.}~\bibnamefont{Marrone}}, \bibnamefont{and}
  \bibinfo{author}{\bibfnamefont{A.}~\bibnamefont{Palazzo}},
  \bibinfo{journal}{Prog. Part. Nucl. Phys.} \textbf{\bibinfo{volume}{102}},
  \bibinfo{pages}{48} (\bibinfo{year}{2018}), \eprint{1804.09678}.

\bibitem[{\citenamefont{Esteban et~al.}(2020)\citenamefont{Esteban,
  Gonzalez-Garcia, Maltoni, Schwetz, and Zhou}}]{Esteban:2020cvm}
\bibinfo{author}{\bibfnamefont{I.}~\bibnamefont{Esteban}},
  \bibinfo{author}{\bibfnamefont{M.~C.} \bibnamefont{Gonzalez-Garcia}},
  \bibinfo{author}{\bibfnamefont{M.}~\bibnamefont{Maltoni}},
  \bibinfo{author}{\bibfnamefont{T.}~\bibnamefont{Schwetz}}, \bibnamefont{and}
  \bibinfo{author}{\bibfnamefont{A.}~\bibnamefont{Zhou}},
  \bibinfo{journal}{JHEP} \textbf{\bibinfo{volume}{09}}, \bibinfo{pages}{178}
  (\bibinfo{year}{2020}), \eprint{2007.14792}.

\bibitem[{\citenamefont{de~Salas et~al.}(2018)\citenamefont{de~Salas, Forero,
  Ternes, Tortola, and Valle}}]{deSalas:2017kay}
\bibinfo{author}{\bibfnamefont{P.~F.} \bibnamefont{de~Salas}},
  \bibinfo{author}{\bibfnamefont{D.~V.} \bibnamefont{Forero}},
  \bibinfo{author}{\bibfnamefont{C.~A.} \bibnamefont{Ternes}},
  \bibinfo{author}{\bibfnamefont{M.}~\bibnamefont{Tortola}}, \bibnamefont{and}
  \bibinfo{author}{\bibfnamefont{J.~W.~F.} \bibnamefont{Valle}},
  \bibinfo{journal}{Phys. Lett. B} \textbf{\bibinfo{volume}{782}},
  \bibinfo{pages}{633} (\bibinfo{year}{2018}), \eprint{1708.01186}.

\bibitem[{\citenamefont{de~Salas et~al.}(2021)\citenamefont{de~Salas, Forero,
  Gariazzo, Mart\'\i{}nez-Mirav\'e, Mena, Ternes, T\'ortola, and
  Valle}}]{deSalas:2020pgw}
\bibinfo{author}{\bibfnamefont{P.~F.} \bibnamefont{de~Salas}},
  \bibinfo{author}{\bibfnamefont{D.~V.} \bibnamefont{Forero}},
  \bibinfo{author}{\bibfnamefont{S.}~\bibnamefont{Gariazzo}},
  \bibinfo{author}{\bibfnamefont{P.}~\bibnamefont{Mart\'\i{}nez-Mirav\'e}},
  \bibinfo{author}{\bibfnamefont{O.}~\bibnamefont{Mena}},
  \bibinfo{author}{\bibfnamefont{C.~A.} \bibnamefont{Ternes}},
  \bibinfo{author}{\bibfnamefont{M.}~\bibnamefont{T\'ortola}},
  \bibnamefont{and} \bibinfo{author}{\bibfnamefont{J.~W.~F.}
  \bibnamefont{Valle}}, \bibinfo{journal}{JHEP} \textbf{\bibinfo{volume}{02}},
  \bibinfo{pages}{071} (\bibinfo{year}{2021}), \eprint{2006.11237}.

\bibitem[{\citenamefont{Davidson et~al.}(1994)\citenamefont{Davidson, Bailey,
  and Campbell}}]{Davidson:1993qk}
\bibinfo{author}{\bibfnamefont{S.}~\bibnamefont{Davidson}},
  \bibinfo{author}{\bibfnamefont{D.~C.} \bibnamefont{Bailey}},
  \bibnamefont{and} \bibinfo{author}{\bibfnamefont{B.~A.}
  \bibnamefont{Campbell}}, \bibinfo{journal}{Z. Phys. C}
  \textbf{\bibinfo{volume}{61}}, \bibinfo{pages}{613} (\bibinfo{year}{1994}),
  \eprint{hep-ph/9309310}.

\bibitem[{\citenamefont{Leurer}(1994{\natexlab{a}})}]{Leurer:1993em}
\bibinfo{author}{\bibfnamefont{M.}~\bibnamefont{Leurer}},
  \bibinfo{journal}{Phys. Rev. D} \textbf{\bibinfo{volume}{49}},
  \bibinfo{pages}{333} (\bibinfo{year}{1994}{\natexlab{a}}),
  \eprint{hep-ph/9309266}.

\bibitem[{\citenamefont{Leurer}(1994{\natexlab{b}})}]{Leurer:1993qx}
\bibinfo{author}{\bibfnamefont{M.}~\bibnamefont{Leurer}},
  \bibinfo{journal}{Phys. Rev. D} \textbf{\bibinfo{volume}{50}},
  \bibinfo{pages}{536} (\bibinfo{year}{1994}{\natexlab{b}}),
  \eprint{hep-ph/9312341}.

\bibitem[{\citenamefont{Carpentier and Davidson}(2010)}]{Carpentier:2010ue}
\bibinfo{author}{\bibfnamefont{M.}~\bibnamefont{Carpentier}} \bibnamefont{and}
  \bibinfo{author}{\bibfnamefont{S.}~\bibnamefont{Davidson}},
  \bibinfo{journal}{Eur. Phys. J. C} \textbf{\bibinfo{volume}{70}},
  \bibinfo{pages}{1071} (\bibinfo{year}{2010}), \eprint{1008.0280}.

\bibitem[{\citenamefont{Bona et~al.}(2008)}]{UTfit:2007eik}
\bibinfo{author}{\bibfnamefont{M.}~\bibnamefont{Bona}} \bibnamefont{et~al.}
  (\bibinfo{collaboration}{UTfit}), \bibinfo{journal}{JHEP}
  \textbf{\bibinfo{volume}{03}}, \bibinfo{pages}{049} (\bibinfo{year}{2008}),
  \eprint{0707.0636}.

\bibitem[{\citenamefont{Abi et~al.}(2021)}]{Muong-2:2021ojo}
\bibinfo{author}{\bibfnamefont{B.}~\bibnamefont{Abi}} \bibnamefont{et~al.}
  (\bibinfo{collaboration}{Muon g-2}), \bibinfo{journal}{Phys. Rev. Lett.}
  \textbf{\bibinfo{volume}{126}}, \bibinfo{pages}{141801}
  (\bibinfo{year}{2021}), \eprint{2104.03281}.

\bibitem[{\citenamefont{Queiroz and Shepherd}(2014)}]{Queiroz:2014zfa}
\bibinfo{author}{\bibfnamefont{F.~S.} \bibnamefont{Queiroz}} \bibnamefont{and}
  \bibinfo{author}{\bibfnamefont{W.}~\bibnamefont{Shepherd}},
  \bibinfo{journal}{Phys. Rev. D} \textbf{\bibinfo{volume}{89}},
  \bibinfo{pages}{095024} (\bibinfo{year}{2014}), \eprint{1403.2309}.

\bibitem[{\citenamefont{Kotila and Iachello}(2012)}]{Kotila:2012zza}
\bibinfo{author}{\bibfnamefont{J.}~\bibnamefont{Kotila}} \bibnamefont{and}
  \bibinfo{author}{\bibfnamefont{F.}~\bibnamefont{Iachello}},
  \bibinfo{journal}{Phys. Rev. C} \textbf{\bibinfo{volume}{85}},
  \bibinfo{pages}{034316} (\bibinfo{year}{2012}), \eprint{1209.5722}.

\bibitem[{\citenamefont{Azzolini et~al.}(2018)}]{CUPID-0:2018rcs}
\bibinfo{author}{\bibfnamefont{O.}~\bibnamefont{Azzolini}} \bibnamefont{et~al.}
  (\bibinfo{collaboration}{CUPID-0}), \bibinfo{journal}{Phys. Rev. Lett.}
  \textbf{\bibinfo{volume}{120}}, \bibinfo{pages}{232502}
  (\bibinfo{year}{2018}), \eprint{1802.07791}.

\bibitem[{\citenamefont{Armengaud et~al.}(2021)}]{CUPID:2020aow}
\bibinfo{author}{\bibfnamefont{E.}~\bibnamefont{Armengaud}}
  \bibnamefont{et~al.} (\bibinfo{collaboration}{CUPID}),
  \bibinfo{journal}{Phys. Rev. Lett.} \textbf{\bibinfo{volume}{126}},
  \bibinfo{pages}{181802} (\bibinfo{year}{2021}), \eprint{2011.13243}.

\bibitem[{\citenamefont{Arnaboldi et~al.}(2003)}]{Arnaboldi:2002te}
\bibinfo{author}{\bibfnamefont{C.}~\bibnamefont{Arnaboldi}}
  \bibnamefont{et~al.}, \bibinfo{journal}{Phys. Lett. B}
  \textbf{\bibinfo{volume}{557}}, \bibinfo{pages}{167} (\bibinfo{year}{2003}),
  \eprint{hep-ex/0211071}.

\bibitem[{\citenamefont{Adams et~al.}(2021)}]{CUORE:2021gpk}
\bibinfo{author}{\bibfnamefont{D.~Q.} \bibnamefont{Adams}} \bibnamefont{et~al.}
  (\bibinfo{collaboration}{CUORE}) (\bibinfo{year}{2021}), \eprint{2104.06906}.

\bibitem[{\citenamefont{Helo et~al.}(2015)\citenamefont{Helo, Hirsch, Ota, and
  Pereira~dos Santos}}]{Helo:2015fba}
\bibinfo{author}{\bibfnamefont{J.~C.} \bibnamefont{Helo}},
  \bibinfo{author}{\bibfnamefont{M.}~\bibnamefont{Hirsch}},
  \bibinfo{author}{\bibfnamefont{T.}~\bibnamefont{Ota}}, \bibnamefont{and}
  \bibinfo{author}{\bibfnamefont{F.~A.} \bibnamefont{Pereira~dos Santos}},
  \bibinfo{journal}{JHEP} \textbf{\bibinfo{volume}{05}}, \bibinfo{pages}{092}
  (\bibinfo{year}{2015}), \eprint{1502.05188}.

\bibitem[{\citenamefont{Mahajan}(2014)}]{Mahajan:2013ixa}
\bibinfo{author}{\bibfnamefont{N.}~\bibnamefont{Mahajan}},
  \bibinfo{journal}{Phys. Rev. Lett.} \textbf{\bibinfo{volume}{112}},
  \bibinfo{pages}{031804} (\bibinfo{year}{2014}), \eprint{1310.1064}.

\bibitem[{\citenamefont{Gonz\'alez et~al.}(2016)\citenamefont{Gonz\'alez,
  Hirsch, and Kovalenko}}]{Gonzalez:2015ady}
\bibinfo{author}{\bibfnamefont{M.}~\bibnamefont{Gonz\'alez}},
  \bibinfo{author}{\bibfnamefont{M.}~\bibnamefont{Hirsch}}, \bibnamefont{and}
  \bibinfo{author}{\bibfnamefont{S.~G.} \bibnamefont{Kovalenko}},
  \bibinfo{journal}{Phys. Rev. D} \textbf{\bibinfo{volume}{93}},
  \bibinfo{pages}{013017} (\bibinfo{year}{2016}), \bibinfo{note}{[Erratum:
  Phys.Rev.D 97, 099907 (2018)]}, \eprint{1511.03945}.

\bibitem[{\citenamefont{Arbel\'aez et~al.}(2016)\citenamefont{Arbel\'aez,
  Gonz\'alez, Hirsch, and Kovalenko}}]{Arbelaez:2016zlt}
\bibinfo{author}{\bibfnamefont{C.}~\bibnamefont{Arbel\'aez}},
  \bibinfo{author}{\bibfnamefont{M.}~\bibnamefont{Gonz\'alez}},
  \bibinfo{author}{\bibfnamefont{M.}~\bibnamefont{Hirsch}}, \bibnamefont{and}
  \bibinfo{author}{\bibfnamefont{S.}~\bibnamefont{Kovalenko}},
  \bibinfo{journal}{Phys. Rev. D} \textbf{\bibinfo{volume}{94}},
  \bibinfo{pages}{096014} (\bibinfo{year}{2016}), \bibinfo{note}{[Erratum:
  Phys.Rev.D 97, 099904 (2018)]}, \eprint{1610.04096}.

\bibitem[{\citenamefont{Arbel\'aez et~al.}(2017)\citenamefont{Arbel\'aez,
  Gonz\'alez, Kovalenko, and Hirsch}}]{Arbelaez:2016uto}
\bibinfo{author}{\bibfnamefont{C.}~\bibnamefont{Arbel\'aez}},
  \bibinfo{author}{\bibfnamefont{M.}~\bibnamefont{Gonz\'alez}},
  \bibinfo{author}{\bibfnamefont{S.}~\bibnamefont{Kovalenko}},
  \bibnamefont{and} \bibinfo{author}{\bibfnamefont{M.}~\bibnamefont{Hirsch}},
  \bibinfo{journal}{Phys. Rev. D} \textbf{\bibinfo{volume}{96}},
  \bibinfo{pages}{015010} (\bibinfo{year}{2017}), \eprint{1611.06095}.

\bibitem[{\citenamefont{Gonz\'alez et~al.}(2018)\citenamefont{Gonz\'alez,
  Hirsch, and Kovalenko}}]{Gonzalez:2017mcg}
\bibinfo{author}{\bibfnamefont{M.}~\bibnamefont{Gonz\'alez}},
  \bibinfo{author}{\bibfnamefont{M.}~\bibnamefont{Hirsch}}, \bibnamefont{and}
  \bibinfo{author}{\bibfnamefont{S.}~\bibnamefont{Kovalenko}},
  \bibinfo{journal}{Phys. Rev. D} \textbf{\bibinfo{volume}{97}},
  \bibinfo{pages}{115005} (\bibinfo{year}{2018}), \eprint{1711.08311}.

\bibitem[{\citenamefont{Ayala et~al.}(2020)\citenamefont{Ayala, Cvetic, and
  Gonzalez}}]{Ayala:2020gtv}
\bibinfo{author}{\bibfnamefont{C.}~\bibnamefont{Ayala}},
  \bibinfo{author}{\bibfnamefont{G.}~\bibnamefont{Cvetic}}, \bibnamefont{and}
  \bibinfo{author}{\bibfnamefont{L.}~\bibnamefont{Gonzalez}},
  \bibinfo{journal}{Phys. Rev. D} \textbf{\bibinfo{volume}{101}},
  \bibinfo{pages}{094003} (\bibinfo{year}{2020}), \eprint{2001.04000}.

\bibitem[{\citenamefont{Lee}(2020)}]{Lee:2020rjh}
\bibinfo{author}{\bibfnamefont{M.~H.} \bibnamefont{Lee}}
  (\bibinfo{collaboration}{AMoRE}), \bibinfo{journal}{JINST}
  \textbf{\bibinfo{volume}{15}}, \bibinfo{pages}{C08010}
  (\bibinfo{year}{2020}), \eprint{2005.05567}.

\end{thebibliography}

\end{document}